%% file: 2024_brir_id_taslp.tex
\documentclass[lettersize,journal]{IEEEtran}
\usepackage{amsmath,amsfonts}
\usepackage{algorithmic}
\usepackage{algorithm}
\usepackage{array}
\usepackage{textcomp}
\usepackage{stfloats}
\usepackage{url}
\usepackage{verbatim}
\usepackage{graphicx}
\usepackage{cite}
\usepackage{subcaption}
\usepackage{bm}
\usepackage{siunitx}

\usepackage{booktabs}
\usepackage{graphicx}

\usepackage{tikz}
\usetikzlibrary{positioning}
\usetikzlibrary {arrows.meta} 
\usetikzlibrary{calc}

\newcommand{\E}[1]{\mathrm{E} \lbrace #1 \rbrace}
\newcommand{\He}{^\mathrm{H}}
\newcommand{\T}{^\mathrm{T}}

\usepackage[absolute,overlay]{textpos}

\begin{document}

\begin{textblock*}{19cm}(1cm,0.5cm) 
\small{\copyright~2024 IEEE. Personal use of this material is permitted. Permission from IEEE must be obtained for all other uses, in any current or future media, including reprinting/republishing this material for advertising or promotional purposes, creating new collective works, for resale or redistribution to servers or lists, or reuse of any copyrighted component of this work in other works.}
\begin{center}
    \textbf{Published in IEEE/ACM TASLP, doi: 10.1109/TASLP.2024.3454964}
\end{center}
\end{textblock*}

\title{Blind Identification of Binaural Room Impulse Responses from Smart Glasses}

\author{Thomas~Deppisch,~\IEEEmembership{Graduate Student Member,~IEEE},
        Nils~Meyer-Kahlen,~\IEEEmembership{Graduate Student Member,~IEEE},
        and~Sebastià~V.~Amengual~Garí,~\IEEEmembership{Member,~IEEE}
\thanks{This research was done during an internship at Reality Labs Research (Meta).}
\thanks{Thomas~Deppisch is with the Chalmers University of Technology, 412 96 Gothenburg, Sweden (e-mail: thomas.deppisch@chalmers.se).}
\thanks{Nils~Meyer-Kahlen is with the Aalto University, 02150 Espoo, Finland (e-mail: nils.meyer-kahlen@aalto.fi).}
\thanks{Sebastià~V.~Amengual~Garí is with Reality Labs Research, Meta, Redmond, WA 98052, USA (e-mail: samengual@meta.com).}
}



\maketitle

\begin{abstract}
Smart glasses are increasingly recognized as a key medium for augmented reality, offering a hands-free platform with integrated microphones and non-ear-occluding loudspeakers to seamlessly mix virtual sound sources into the real-world acoustic scene. To convincingly integrate virtual sound sources, the room acoustic rendering of the virtual sources must match the real-world acoustics. Information about a user's acoustic environment however is typically not available. This work uses a microphone array in a pair of smart glasses to blindly identify binaural room impulse responses (BRIRs) from a few seconds of speech in the real-world environment. The proposed method uses dereverberation and beamforming to generate a pseudo reference signal that is used by a multichannel Wiener filter to estimate room impulse responses which are then converted to BRIRs. The multichannel room impulse responses can be used to estimate room acoustic parameters which is shown to outperform baseline algorithms in the estimation of reverberation time and direct-to-reverberant energy ratio. Results from a listening experiment further indicate that the estimated BRIRs often reproduce the real-world room acoustics perceptually more convincingly than measured BRIRs from other rooms of similar size. 
\end{abstract}


\section{Introduction}\label{sec:introduction}
\begin{figure}[t]
  \centering
  \includegraphics[trim=0cm 1cm 0cm 5cm,clip,width=0.7\columnwidth]{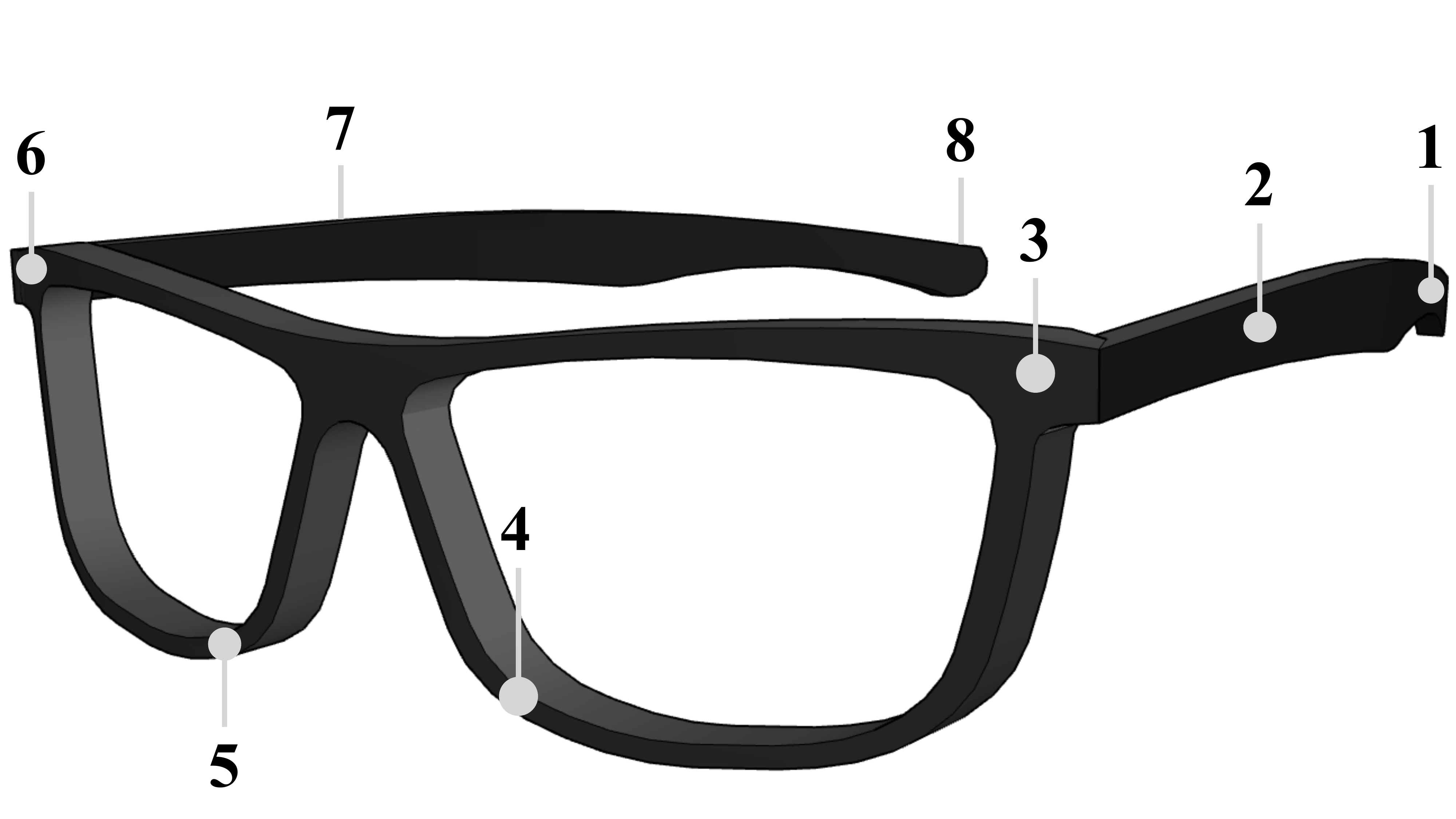}
  \caption{Microphone positions on a pair of glasses as used in this study.}
  \label{fig:glassesWithMicPos}
\end{figure}

\IEEEPARstart{A}{udio} for augmented reality (AR) aims to augment the real world with virtual sound sources that realistically blend into the acoustic scene. As part of such a system, the room acoustic rendering of virtual sources must match the acoustics of the room in which the user is located~\cite{Neidhardt2022b, Gari2022}. AR applications are often enabled by head-worn devices such as head-mounted displays (HMDs) or smart glasses. In the present contribution, we propose a method that addresses the acoustic matching problem under realistic acoustic conditions: we use a microphone array that is integrated into a pair of smart glasses as illustrated in Fig.~\ref{fig:glassesWithMicPos} to estimate BRIRs in noisy, real-world environments. 

BRIRs represent the linear, time-invariant acoustic transfer path between a sound source and the sound pressure at a listener's eardrum. BRIRs thus include the acoustic properties of the environment and the direction-dependent influence of the listener's head, torso, and outer ear, as captured by a set of head-related impulse responses (HRIRs). Although HRIRs are highly individual to the listener's morphology~\cite{Moller1995a}, scalable and widespread personalization of HRIRs is not currently feasible, and binaural rendering of virtual sound sources for a general audience therefore often replaces the individual HRIRs with a generic set of HRIRs from a dummy head. 
\IEEEpubidadjcol

Yet even binaural rendering based on BRIRs measured with a dummy head is still infeasible in most practical AR applications, for two main reasons: (i), a large set of dummy head BRIRs is required to facilitate head rotations during the rendering and, (ii), dedicated acoustic measurements in the target environment are infeasible in consumer AR applications. 

The first challenge can be overcome by employing an array that captures room impulse responses (RIRs) with multiple microphones to characterize the directional properties of the acoustic environment. Using array processing techniques, the array RIRs are transformed into BRIRs for any given head rotation by combining them with an anechoically measured set of generic HRIRs~\cite{AmengualGari2020,Deppisch2021a}. Perceptually plausible rendering from array RIRs, i.e., convincing rendering when no explicit external reference is provided, has been achieved with such a method~\cite{AmengualGari2020}.

The second challenge can be overcome by blind estimation of the array RIRs from sounds that naturally occur in the user's environment like own speech or the speech of other people. While established signal processing and machine-learning methods for blind RIR estimation exist in the literature, they are typically not designed and validated for the estimation of multichannel RIRs supporting the full audible frequency range under realistic acoustic conditions and may fail to converge in such cases~\cite{Deppisch2024}. 


The herein proposed method extends the work presented in~\cite{Meyer-Kahlen2022c, Deppisch2024}. It uses a pseudo reference signal obtained by beamforming and dereverberation from a few seconds of captured speech and then identifies a multichannel RIR which forms the basis for the subsequent binaural rendering. New contributions in this work are the extension of the signal model to consider noise, the use of the wearer's own voice for the estimation, a resynthesis approach for the estimated RIRs and binaural rendering to obtain BRIRs, an extensive evaluation using a data set of measured RIRs from a pair of glasses, the analysis of the robustness of the method, 
and the perceptual evaluation of the whole processing chain.

\section{Background}\label{sec:background}
Several approaches for the room acoustic matching of virtual sound sources with real sources have been explored in the literature. One option is to estimate RIRs from running signals and use convolution to match the acoustics of the virtual source signal with the estimated real-world acoustics. Most of the proposed methods solve the blind multichannel identification task by exploiting cross-relations between channels, such as~\cite{Xu1995, Huang2003, Haque2008, Jo2021}. However, the authors showed in~\cite{Deppisch2024} that cross-relation-based methods fail to converge if acoustic RIRs of realistic lengths (multiple hundreds of milliseconds) are estimated. Thus, new methods that converge quicker have been proposed~\cite{Perez-Lopez2020, Meyer-Kahlen2022c, Deppisch2024}. They transform the blind identification task into a non-blind one by applying dereverberation and/or beamforming to estimate a pseudo reference signal. The transfer function between the pseudo reference signal and the array microphone signals then serves as a room transfer function estimate and its time-domain counterpart as RIR. Promising results have been obtained for reverberation time (RT) estimation using simulated omnidirectional RIRs~\cite{Perez-Lopez2020} and multichannel RIRs~\cite{Meyer-Kahlen2022c} in the spherical harmonics (SH) domain. The generalized approach from~\cite{Deppisch2024} does not require SH-domain processing. So far, the method has been validated with simulated and measured multichannel RIRs using spherical and circular arrays with regularly distributed microphones. 

Recently, the RIR estimation task has also been approached using deep-learning (DL) methods. For example, \cite{Steinmetz2021a} uses a custom architecture with an encoder network and a decoder that models the response using shaped noise. \cite{Lee2023a} built on this architecture, but allowed for multiple sources in the room by estimating one generic response per room. Other recent approaches use generative networks~\cite{Liao2023, Ratnarajah2023, Lee2023}. So far, the DL-based methods only consider the single-channel problem and most often work with sampling rates that do not support the entire audible frequency range. Most of the DL-based evaluations only consider numerical metrics and do not include a perceptual evaluation. An exception is~\cite{Steinmetz2021a} which solved the single-channel task for the full audible bandwidth and delivered promising parameter estimation as well as perceptual results. It is noted therein that other DL-based methods often generate audible artifacts. The extension of the DL methods to support multichannel RIR estimation and binaural rendering would require training data from a large number of realistic acoustic environments for a specific microphone array geometry and would thus demand extensive measurement or simulation efforts. 

An alternative to the full RIR estimation is to only estimate a set of room acoustic parameters and resynthesize a room response from them, turning the task into a blind parameter estimation problem. Blind parameter estimation has been a topic of research for many years; a comparison of different methods was made through the Acoustic Characterization of Environments (ACE) challenge~\cite{Eaton2016}. 
Since then, several machine-learning-based algorithms for parameter estimation have been proposed, such as~\cite{Gamper2018, Gotz2023}. So far, the motivation for parameter estimation was not using the parameters to render sound but to inform other algorithms, for example for speech enhancement or recognition, in order to improve their performance. For that reason, the algorithms do not contain stages for resynthesizing a response.

Thus far, it is unknown whether full RIR estimation or parameter estimation is the most promising approach for AR. 
Computing parameters from RIR estimates is beneficial for performance assessment as a basic set of parameters such as RT and direct-to-reverberant energy ratio (DRR) are easier to interpret and correlate better to certain attributes of room acoustic perception than technical measures such as projection misalignment. Therefore, we also use parameters to evaluate our RIR estimation and compare the results to the best-performing algorithms from the ACE challenge. We believe, however, that estimating the full RIR offers a clear advantage over estimating parameters alone as any parameter can be derived from a full multichannel RIR estimate and it is still unclear which exact set of parameters needs to be reproduced to achieve perceptually adequate, i.e., plausible~\cite{Lindau2012} or transfer-plausible~\cite{Wirler2020}, rendering.

\section{RIR Estimation}\label{sec:rir_estimation}
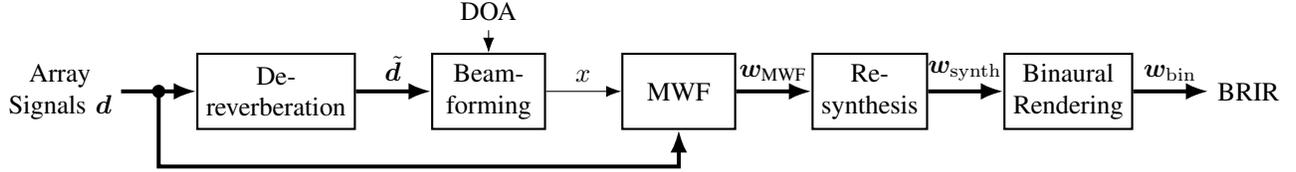
\begin{figure*}
\centering
\begin{tikzpicture}[
squarednode/.style={rectangle, draw=black, thick, minimum height=1cm, minimum width=1.5cm, align=center},
node distance=10mm
]
\node[align=center]                   (start)                                  {Array\\Signals $\bm d$};
\node[squarednode]      (derev)         [right=of start]         {De-\\reverberation};
\node[squarednode]      (beamf)         [right=of derev]         {Beam-\\forming};
\node[squarednode]      (mwf)           [right=of beamf]         {MWF};
\node[node distance=3mm]      (doa)           [above=of beamf]         {DOA};
\node[squarednode]      (resynth)       [right=of mwf]      {Re-\\synthesis};
\node[squarednode]      (binaural)       [right=of resynth]      {Binaural\\Rendering};
\node                   (end)           [right=of binaural, align=center]           {BRIR};
\coordinate             (fork1) at ($(start.east)!0.5!(derev.west)$); 

\draw[-{Latex[length=3mm]}, line width=1.5pt] (start.east) -- (derev.west);
\draw[-{Latex[length=3mm]}, line width=1.5pt] (derev.east) -- (beamf.west) node[above, pos=0.49, align=center] {$\tilde{\bm d}$};
\draw[-{Latex[length=2mm]}] (beamf.east) -- (mwf.west) node[above, pos=0.49, align=center] {$x$};
\draw[-{Latex[length=2mm]}] (doa.south) -- (beamf.north);
\draw[-{Latex[length=3mm]}, line width=1.5pt] (mwf.east) -- (resynth.west) node[above, pos=0.49, align=center] {$\bm w_{\text{MWF}}$};
\draw[-{Latex[length=3mm]}, line width=1.5pt] (resynth.east) -- (binaural.west) node[above, pos=0.49, align=center] {$\bm w_\mathrm{synth}$};
\draw[-{Latex[length=3mm]}, line width=1.5pt] (binaural.east) -- (end.west) node[above, pos=0.49, align=center] {$\bm w_\mathrm{bin}$};
\draw[-{Latex[length=3mm]}, line width=1.5pt] (fork1) -- ++(0,-1) -- ($(mwf) + (0,-1)$) node[above, pos=0.8] {} -- (mwf.south);
\fill (fork1) circle (2.5pt);
\end{tikzpicture}
\caption{The proposed processing estimates a binaural room impulse response (BRIR) $\bm w_\mathrm{bin}$ from the array signals $\bm d$ via the pseudo reference signal $x$. Bold lines represent multichannel signals.}
\label{fig:signalFlow}
\end{figure*}

\subsection{Signal Model}
Consider a person speaking in a room and the speech being picked up by a microphone array in a pair of smart glasses that is either worn by the speaker or by another person in the room.
The array signals can then be described in discrete time $n$ as the convolution of the speech signal $s(n)$ and the multichannel RIR $\bm h(n)$ from the speaker's mouth to each microphone plus additive noise $\bm v(n)$,
\begin{equation}\label{eq:basicSignalModel}
    \bm d(n) = \bm h(n) * s(n) + \bm v(n) \, .
\end{equation}
$M$ is the number of microphones in the array and $\bm d(n)$, $\bm h(n)$ and $\bm v(n)$ are length-$M$ vectors. Lowercase and uppercase bold letters in this contribution denote vectors and matrices, respectively.
The herein proposed method estimates the multichannel RIR $\bm h(n)$ by utilizing an estimate of the speech signal $s(n)$ which is referred to as pseudo reference signal $x(n)$. By exploiting the pseudo reference signal, the method converts the blind identification task into a non-blind one so that the RIR estimate can be obtained using a multichannel Wiener filter (MWF).

For the estimation of the pseudo reference signal, it is beneficial to re-express the RIR $\bm h(n)$ as a sum of the individual contributions of the direct sound path $\bm h_\mathrm{d}(n)$, the early reflections $\bm h_\mathrm{e}(n)$, and the late reverberation $\bm h_\mathrm{l}(n)$, $\bm h(n) = \bm h_\mathrm{d}(n) + \bm h_\mathrm{e}(n) + \bm h_\mathrm{l}(n)$.
The signal model from~\eqref{eq:basicSignalModel} is then re-expressed in the frequency domain,
\begin{equation}\label{eq:fullSignalModel}
    \bm d(\omega) = \underbrace{\bm h_\mathrm{d}(\omega) s(\omega)}_{\text{Direct Sound}}
    + \underbrace{\bm h_\mathrm{e}(\omega) s(\omega)}_{\text{Early Reflections}} + \underbrace{\bm h_\mathrm{l}(\omega) s(\omega)}_{\text{Late Reverb}} + \underbrace{\bm v(\omega)}_{\text{Noise}} \, ,
\end{equation}
where $\omega$ is the angular frequency.
For notational convenience, we distinguish between time- and frequency-domain representations of the signals solely by exchanging the dependent variable, i.e., $\bm d(\omega)$ denotes the discrete Fourier transform of $\bm d(n)$. 

The proposed processing aims to recover an estimate of the signal $s(\omega)$ as pseudo reference signal while only having access to $\bm d(\omega)$. This is achieved by reducing the influence of the late reverberation using dereverberation and employing a beamformer with distortionless-response constraint to extract the pseudo reference undistorted while minimizing the influence of the early reflections and noise. 
Once the pseudo reference signal has been obtained, a multichannel RIR is estimated via an MWF. 
As detailed in Sec.~\ref{sec:rendering}, the late part of the RIR estimate is then resynthesized using filtered noise and is converted to a binaural response. 
The full processing chain is illustrated in Fig.~\ref{fig:signalFlow}.

\subsection{Dereverberation}
The generalized weighted prediction error (GWPE) method~\cite{Yoshioka2012} is chosen for the blind multichannel dereverberation as it provides a high dereverberation performance while avoiding signal distortions and achieving a high perceptual signal quality~\cite{Neo2021a}. The GWPE method is especially well suited for the given problem as it produces the same number of output channels as input channels, does not require knowledge of the number of sources, and preserves time differences between channels. It is integrated into a short-time Fourier transform (STFT) processing framework where each bin is interpreted as time-variant complex-valued subband signal with time index $\eta$. The method estimates a multichannel prediction filter matrix $\bm G_b(\eta)$ with filters of length $K_b$ in each subband $b$ that minimizes the temporal signal correlation after a prediction delay $\Delta$ to obtain the dereverberated subband signals
\begin{equation}
    \tilde{\bm d}_b(\eta) = \bm d_b(\eta) - \sum_{\tau=\Delta}^{\Delta+K_b-1} \bm G_b\He(\tau)\, \bm d_b(\eta - \tau) \, ,
\end{equation}
where $(\cdot)\He$ is the Hermitian transpose.
The prediction delay $\Delta$ is typically chosen in the range of tens of milliseconds so that the short-term autocorrelation of speech is not affected by the filter. Thus, only the late reverberation in~\eqref{eq:fullSignalModel} with a delay greater than the prediction delay is suppressed by the GWPE algorithm. Note that the GWPE method exploits multichannel information and thus shows better dereverberation performance when more channels are available. For this reason, the dereverberation is applied before the beamformer, see Fig.~\ref{fig:signalFlow}. In this work, we assume time-invariant conditions and thus non-adaptive dereverberation is sufficient. However, the GWPE can be replaced by adaptive dereverberation algorithms with similar properties such as~\cite{Yoshioka2013, Braun2016} to support adaptive processing. 

\subsection{Beamforming}
A minimum variance distortionless response (MVDR) beamformer~\cite[Ch.~6.2.1]{VanTrees2002} is employed to extract the signal $s(\omega)$ from the direct sound component while suppressing the early reflections and the noise described by the signal model in~\eqref{eq:fullSignalModel}. 
With the noise power spectral density (PSD) matrix $\bm P_\mathrm{noise}(\omega) = \E{ \bm v(\omega) \bm v\He(\omega)}$ and the steering vector $\bm a(\omega)$, the MVDR beamformer weights $\bm w_{\text{BF}}(\omega)$ are obtained as
\begin{equation}\label{eq:MVDR_beamformer}
    \bm w_{\text{BF}}(\omega) = \frac{\bm P_\mathrm{noise}^{-1}(\omega) \bm a(\omega)}{\bm a\He(\omega) \bm P_\mathrm{noise}^{-1}(\omega) \bm a(\omega)} \, .
\end{equation}
The noise PSD matrix is typically estimated during speech pauses by assuming ergodic signals and replacing the expectation $\E{\cdot}$ with a temporal average.
The pseudo reference signal $x(\omega)$ is then obtained by applying the beamformer to the dereverberated array signals $\tilde{\bm d}(\omega)$,
\begin{equation}
    x(\omega) = \bm w_{\text{BF}}\He (\omega) \tilde{\bm d}(\omega) \, .
\end{equation}
In practice, the processing is applied blockwise using the STFT.
We assume access to measured anechoic array transfer functions (ATFs) for a dense grid of directions. In the processing, the ATF from the source direction is used as steering vector $\bm a(\omega)$. Thus, knowledge of the source direction of arrival (DOA) is required, and in~\cite{Deppisch2024}, the multiple signal classification (MUSIC) algorithm~\cite{Schmidt1986} was successfully used as DOA estimator for the proposed method. In the present contribution, we provide the algorithm with the true DOAs and investigate the influence of a DOA mismatch separately in Sec.~\ref{sec:robustness}.

\subsection{Transfer Function Estimation}
With the pseudo reference signal $x(\omega)$, the blind estimation problem is transformed into a non-blind one and conventional system identification methods can be applied to obtain the RIR estimate. As in~\cite{Deppisch2024}, we propose to use an MWF to minimize the mean squared error (MSE) between the filtered pseudo reference ${\bm y(\omega) = \bm w_{\text{MWF}}(\omega) x(\omega)}$ and the array signals $\bm d(\omega)$,
\begin{equation}
    \bm w_{\text{MWF}}(\omega) = \frac{1}{\Phi_{xx}(\omega) + \delta} \bm \Phi_{xd}(\omega)\, ,
\end{equation}
where ${\Phi_{xx}(\omega) = \E{x^*(\omega) x(\omega)}}$ is the PSD of the pseudo reference signal, ${\bm \Phi_{xd}(\omega) = \E{x^*(\omega) \bm d(\omega)}}$ is the cross spectral density (CSD) vector, $(\cdot)^*$ denotes the complex conjugate, and $\delta$ is a small regularization constant.
The multichannel RIR estimate is obtained as the time-domain counterpart of the transfer function estimate.

In this work, we perform batch processing using the MWF, assuming static sources and time-invariant conditions. 
Equivalently, a recursive least squares (RLS) filter can be employed as in~\cite{Meyer-Kahlen2022c} if time-variant conditions are assumed.\footnote{The MWF and RLS are equivalent if the forgetting factor of the RLS is set to one, the RLS is initialized with zeros, and the PSD and CSD estimates are obtained in the same way.}
The herein proposed implementation of the method prioritizes optimal performance over computational complexity. While recognizing the importance of computational efficiency for real-time, on-device applications, we reserve the development of an adaptive online algorithm for future research. Currently, dereverberation is identified as the most computationally intensive processing block, and the most important change to decrease complexity will be to exchange the dereverberation module for a more efficient online algorithm.

\section{Rendering}\label{sec:rendering}

\subsection{Ringing Artifacts}
If a virtual source signal is to be rendered by convolution with a BRIR estimate, it is essential for the estimate to not contain audible artifacts. However, the obtained RIR estimates in some cases exhibit narrowband ringing artifacts due to narrowband spectral nulls in the pseudo reference signal that are more significant than corresponding nulls in the array signals. Such nulls are created by interfering sound waves as described by the room and the array transfer functions and may even be reinforced by the dereverberation. In other words, the transfer function of the reference and the array may share common zeros, which is a violation of an identifiability condition known from blind identification via channel cross-relations~\cite{Xu1995}. 

\begin{figure}[t]
  \centering
  \includegraphics[trim=0.5cm 1.9cm 1cm 0cm,clip,width=\columnwidth]{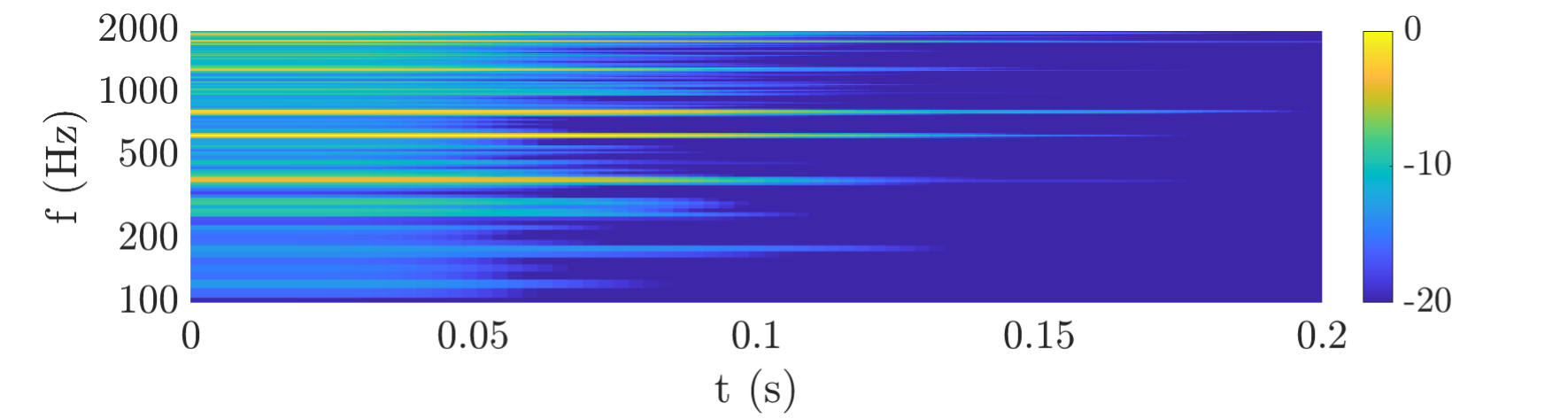}
  \put(-220,5){\footnotesize{\textcolor{white}{Estimated RIR}}}\\
  \includegraphics[trim=0.5cm 0cm 1cm 0cm,clip,width=\columnwidth]{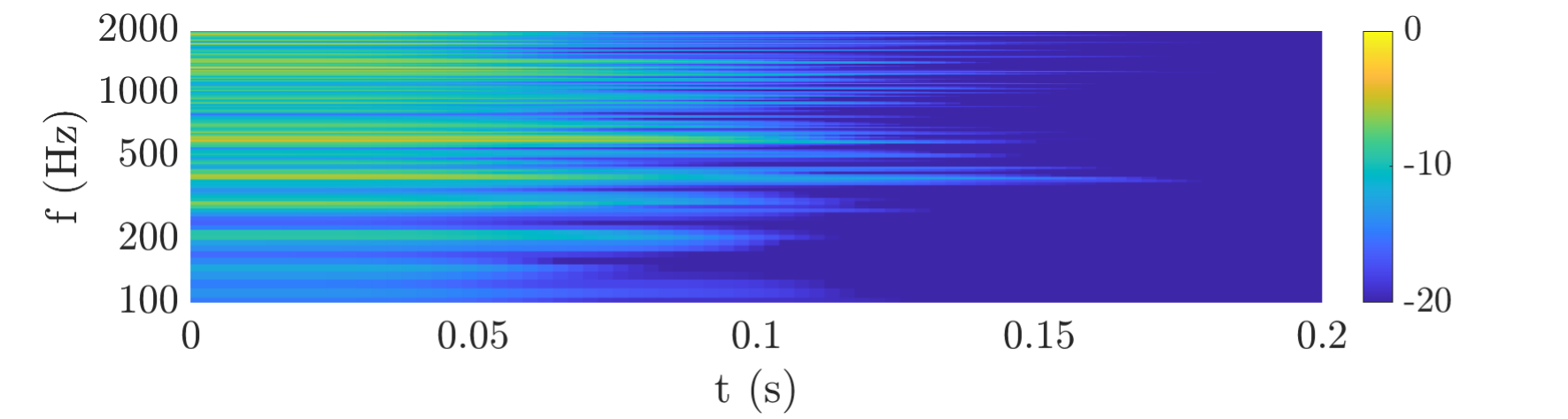}
  \put(-220,22){\footnotesize{\textcolor{white}{Resynthesized RIR}}}
  \caption{The EDR of a single channel of an RIR estimate (top) contains large narrowband energies that may be audible as ringing artifacts. The ringing is suppressed in the EDR of the resynthesized RIR (bottom).}
  \label{fig:edr}
\end{figure}

An example of this is given in Fig.~\ref{fig:edr}, which shows the energy decay relief (EDR) of one channel of an RIR estimate using the proposed processing. The EDR is calculated as the frequency-dependent, reverse-integrated energy of the RIR~\cite{Jot1992}. The EDR is typically normalized to \SI{0}{dB} at each frequency to allow for the estimation of reverberation times but we omit the normalization to illustrate the relative energy content. The estimated RIR shows large, slowly decaying energy contributions in some narrow bands that manifest themselves as audible ringing when a speech sample is convolved with it.\footnote{Audio examples are provided at \texttt{\url{https://facebookresearch.github.io/GlassesRoomID/}}.} Note that the ringing has a strong influence on the perceived quality but only little impact on metrics using a coarser frequency resolution like octave bands. This can be observed in the results from Sec.~\ref{sec:objective} that were obtained using the raw estimates without the resynthesis.

\subsection{RIR Resynthesis}\label{sec:rirResynth}
As the ringing only occurs in narrow frequency bands, we propose to resynthesize the RIR estimates in octave bands using filtered noise. Synthesized RIRs based on filtered noise have been shown to be perceptually convincing~\cite{Traer2016} and by resynthesizing the RIRs with a coarse, octave-band frequency resolution, narrowband inaccuracies such as the ringing are prevented in the resynthesized response.

\begin{figure}[t]
  \centering
  \includegraphics[trim=1cm 1cm 1cm 1cm,clip,width=\columnwidth]{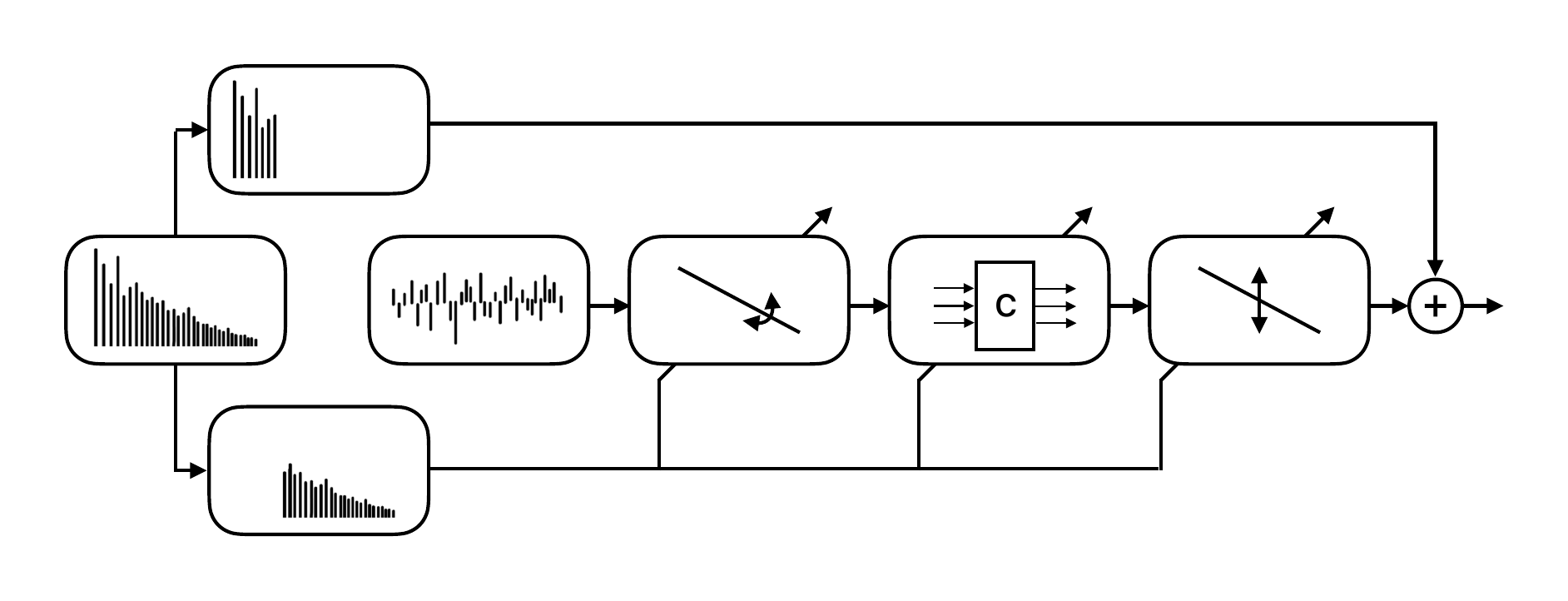}
  \put(-219,-4){\scriptsize{Late Part}}
  \put(-220,55){\scriptsize{Early Part}}
  \put(-186,26){\scriptsize{Noise}}
  \put(-142,26){\scriptsize{Decay}}
  \put(-105,26){\scriptsize{Covariance}}
  \put(-55,26){\scriptsize{Energy}}
  \caption{The RIR estimate is resynthesized by replacing the late part of the estimate with filtered noise. The noise-filtering process matches the exponential decay, the covariance matrix, and the early-to-late energy ratio in octave bands.}
  \label{fig:resynth_block_diagram}
\end{figure}
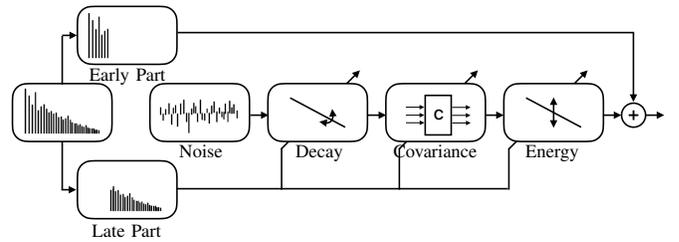

The synthesis of late reverberation tails using noise is well known from the literature~\cite{Traer2016,Valimaki2017a, Porschmann2017, Arend2021a}. Typically, the methods either match the noise EDR in time-frequency blocks or match the decay and the early-to-late energy ratio. We follow the latter approach as it automatically removes any noise floor in the RIR estimates. We additionally perform covariance matching to preserve inter-channel relations. 
The full procedure is illustrated in Fig.~\ref{fig:resynth_block_diagram} and the resulting EDR is shown in Fig.~\ref{fig:edr}. Note that the proposed method does not support the detection and synthesis of multi-slope decays but could be extended using~\cite{Hold2022,Gotz2022a}.

In the first step, the RIR estimate is decomposed into octave bands using a filter bank and divided into an early part and a late part at time $\tau_\mathrm{split}$. Ideally, $\tau_\mathrm{split}$ should be greater than the time of arrival of perceptually salient early reflections and small enough to suppress the ringing. It was however found through informal listening that the impact of the exact transition time is minimal and it was thus set to $\tau_\mathrm{split} = \SI{20}{ms}$ for all resynthesized RIRs in this study. Next, $M$ independent realizations of normally distributed noise of the same length as the late part of the RIR estimate are generated and processed by the filter bank. An exponential decay slope according to the reverberation time $\tau_{m,b}$ of the estimate is applied to the generated noise $\nu_{m,b}(n)$, 
\begin{equation}
    \nu_{m,b}^\mathrm{decay}(n) = 10^{\frac{-60\, n}{20 \, \tau_{m,b} f_\mathrm{s}}} \,  \nu_{m,b}(n) \, ,
\end{equation}
where $m$ denotes the microphone channel index, $b$ denotes the band index of the filter bank, and $f_\mathrm{s}$ is the sampling frequency. The spatial covariance of the generated decaying noise is matched to the covariance of the estimate $\bm C_\mathrm{target}$ by exploiting the eigendecomposition $\bm C_\mathrm{target} = \bm V \bm D \bm V\T$~\cite{Habets2008},
\begin{equation}
    \bm \nu_b^\mathrm{cov}(n) =  \bm V \sqrt{\bm D} \, \bm \nu_b^\mathrm{decay}(n) \, .
\end{equation}
Here, the decaying noise of all $M$ microphone channels are stacked in the vector ${\bm \nu_b^\mathrm{decay}(n) = [\nu_{1,b}^\mathrm{decay}(n), \, \dots, \, \nu_{M,b}^\mathrm{decay}(n)]\T}$, the diagonal matrix $\bm D$ contains the eigenvalues $\bm C_\mathrm{target}$, and the matrix $\bm V$ contains the orthonormal eigenvectors.

\begin{figure}[t!]
  \centering
  \includegraphics[trim=0cm 1.9cm 0cm 0cm,clip,width=\columnwidth]{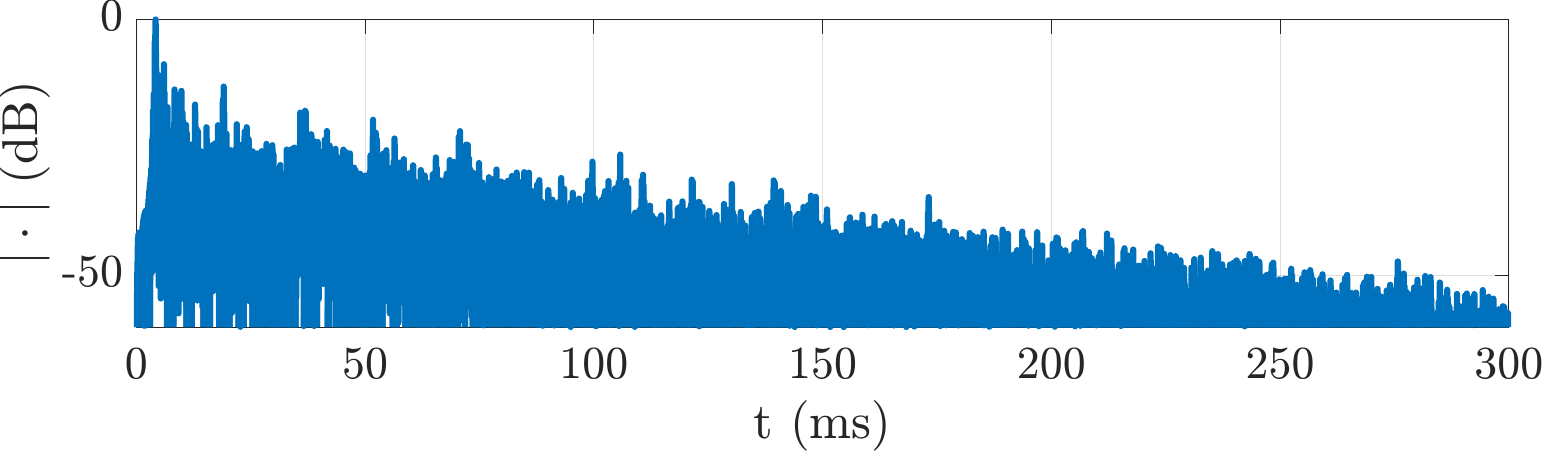}
  \put(-70,43){\footnotesize{Ground Truth RIR}}\\
  \includegraphics[trim=0cm 1.9cm 0cm 0cm,clip,width=\columnwidth]{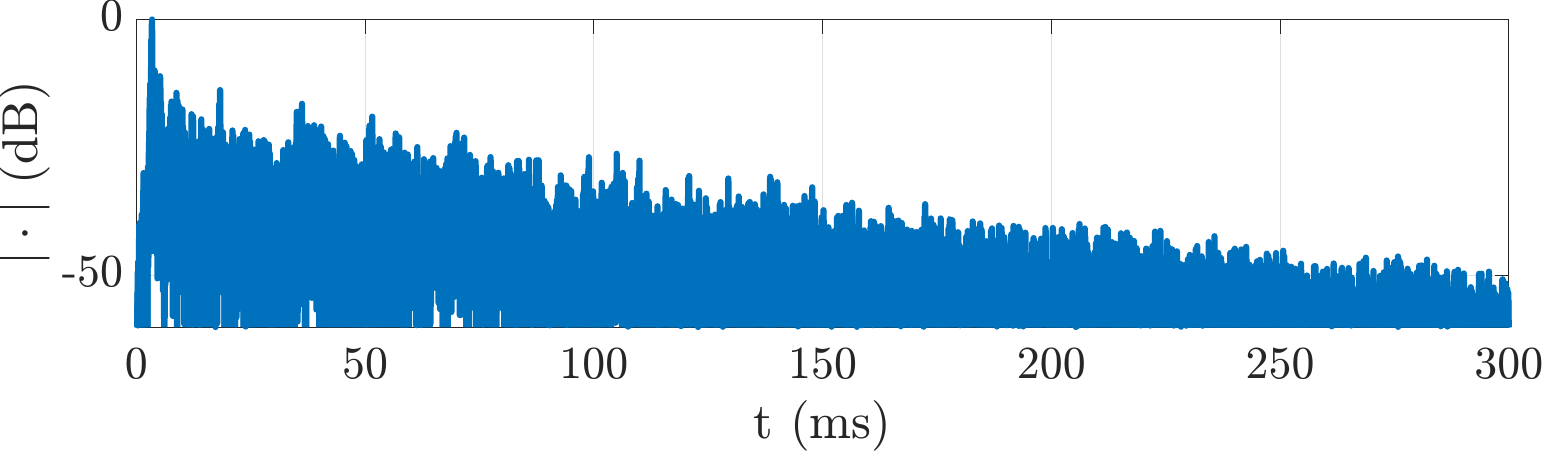}
  \put(-58,43){\footnotesize{Estimated RIR}}\\
  \includegraphics[width=\columnwidth]{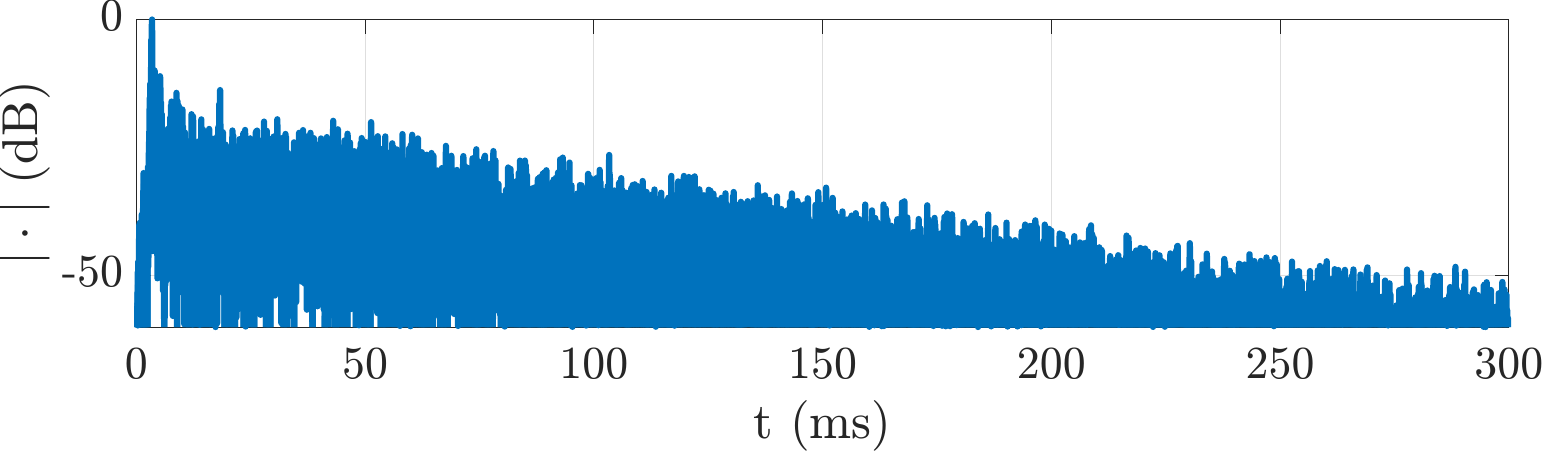}
  \put(-72,61){\footnotesize{Resynthesized RIR}}
  \caption{One channel of a measured RIR (top), a corresponding RIR estimate (center), and the resynthesized RIR (bottom).}
  \label{fig:rirsTimeDomain}
\end{figure}

In the third step, the energy of the covariance-matched noise $\nu_{m,b}^\mathrm{cov}(n)$ is matched to the energy of the late part of the estimated RIR $w_{m,b}^\mathrm{MWF,late}(n)$,
\begin{equation}
    \nu_{m,b}^\mathrm{en}(n) =  \sqrt{\frac{\sum_n (w_{m,b}^\mathrm{MWF,late}(n))^2}{\sum_n (\nu_{m,b}^{\mathrm{cov}}(n))^2}} \, \nu_{m,b}^\mathrm{cov}(n) \, .
\end{equation}
The energy matching ensures that the early-to-late energy ratio of the RIR estimate is preserved in the resynthesized RIR. Lastly, the resynthesized RIR estimate is obtained by concatenating the early part of the RIR estimate and the filtered noise in the time domain and stacking the results for all microphone channels in the vector $\bm w_\mathrm{synth}(n)$. 
A measured RIR from a meeting room, a corresponding estimate from the proposed method, and its resynthesized version are shown in Fig.~\ref{fig:rirsTimeDomain}.

The proposed resynthesis is designed for estimates obtained from far-field speech. RIR estimates can also be generated from the speech of the user wearing the smart glasses which is discussed in Sec.~\ref{sec:own_voice}. The synthesis of far-field RIRs from own-speech estimates requires additional processing to assign a meaningful distance and direction and is thus left for future work.

\subsection{Binaural Rendering}\label{sec:binauralRendering}
To render a virtual sound source via headphones, its anechoic source signal $s(\omega)$ is filtered with a BRIR estimate $w_\mathrm{bin}^\mathrm{\lbrace l,r \rbrace}(\omega)$,
\begin{equation}\label{eq:brirConvolution}
    y_\mathrm{virt}^\mathrm{\lbrace l,r \rbrace}(\omega) = w_\mathrm{bin}^\mathrm{\lbrace l,r \rbrace}(\omega) \, s(\omega).
\end{equation}
The rendering is performed for the left- and right-ear signals as indicated by the superscript $\mathrm{\lbrace l,r \rbrace}$. The remainder of this section describes how the BRIR estimate $w_\mathrm{bin}^\mathrm{\lbrace l,r \rbrace}(\omega)$ is computed from the multichannel RIR estimate $\bm w_\mathrm{synth}(\omega)$.

The binaural rendering of multichannel array signals is an actively researched problem. Due to the simplifications of the RIR estimate that were introduced by replacing the late part of the RIR with filtered noise, we only consider signal-independent, non-parametric rendering methods, i.e., methods that do not require the estimation of parameters like reflection directions. Depending on the specific assumptions on the microphone array and sound field, three non-parametric binaural rendering methods are considered state-of-the-art: magnitude-least-squares-optimal rendering (magLS)~\cite{Zaunschirm2018,Schorkhuber2018, Deppisch2021a}, beamforming-based binaural reproduction (BFBR)~\cite{Duraiswami2005,Ifergan2022}, and binaural signal matching (BSM)~\cite{Madmoni2020,Madmoni2021}. The most recent iterations of these methods often combine ideas of the approaches, e.g., in~\cite{Lubeck2022} BSM was implemented with magnitude optimization at high frequencies and \cite{Madmoni2020} combines BFBR and BSM. 
In the following, the end-to-end magnitude least squares method (eMagLS)~\cite{Deppisch2021a} is used as it does not require additional assumptions, specific microphone arrays or the choice of rendering parameters, and was recently shown to outperform other non-parametric methods in a listening experiment~\cite{Stahl2023a}.

The eMagLS method transforms the multichannel RIR into a BRIR by applying optimal rendering filters $\bm w_\mathrm{MLS}(\omega)$ and summing over all microphone channels,
\begin{equation}\label{eq:binauralRendering}
    w_\mathrm{bin}^\mathrm{\lbrace l,r \rbrace}(\omega) = (\bm w_\mathrm{MLS}^\mathrm{\lbrace l,r \rbrace}(\omega))\He \, \bm w_\mathrm{synth}(\omega) \, .
\end{equation}
The eMagLS rendering filters ensure a minimum \mbox{(magnitude-)} least-squares rendering error, i.e., they minimize the difference between a head-related transfer function (HRTF, the frequency-domain counterpart of the HRIR) and a filtered array signal that is generated by a plane wave from the same direction as the HRTF for a large set of directions.
While in the original publication, the eMagLS method was developed using analytically derived ATFs for spherical microphone arrays, the method can be used in the same way with arbitrary arrays using measured ATFs which has also recently been explored in~\cite{McCormack2023}. We use measured ATFs to calculate the rendering filters without utilizing a spherical harmonics decomposition, which was termed \textit{eMagLS2} in~\cite{Deppisch2021a}.
The eMagLS filters are least-squares-optimal at low frequencies and minimize the least-squares error of the magnitude at high frequencies, favoring the accurate rendering of magnitudes of binaural cues over their phase. If a dynamic binaural reproduction including compensation for a user's head rotation is required, the rendering filters can be computed with a rotated HRTF set as shown in~\cite{McCormack2023}. In this case, different filters $\bm w_\mathrm{MLS}^\mathrm{\lbrace l,r \rbrace}(\omega)$ need to be selected in~\eqref{eq:binauralRendering} depending on the head rotation. 

\section{Objective Evaluation}\label{sec:objective}
\subsection{Data Set}\label{sec:evalDataSet}
\begin{figure}[t]
  \centering
  \includegraphics[width=\columnwidth]{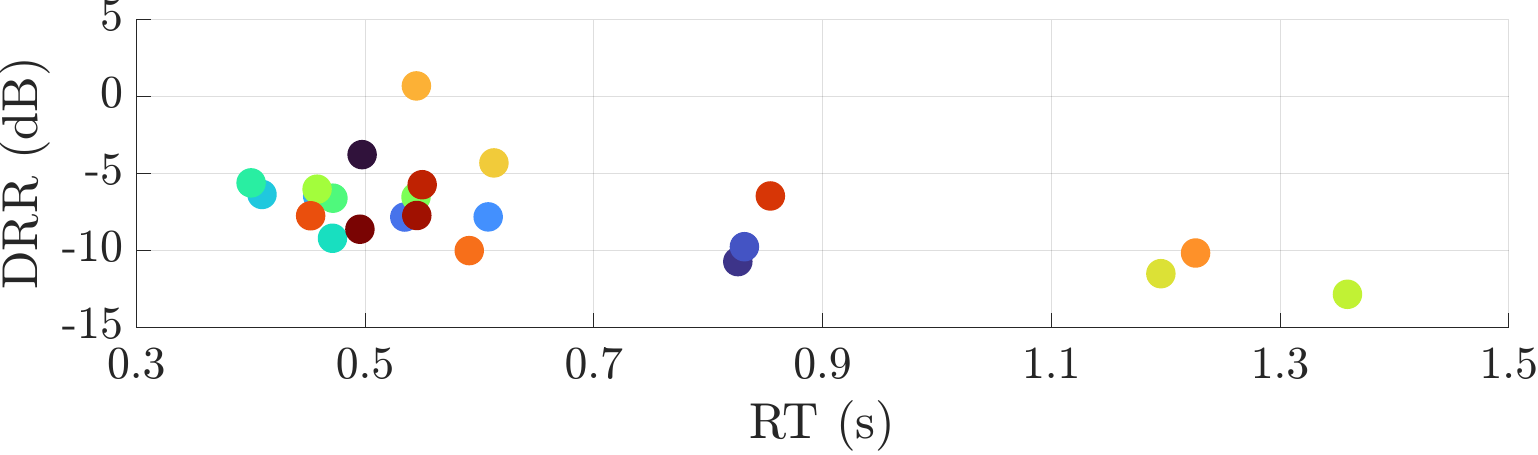}
  \caption{Average RTs and DRRs of the measurements from the 23 rooms in the data set.}
  \label{fig:dataSetRts}
\end{figure}

We evaluate the proposed method using a data set of 23~rooms. In each room, RIRs were measured using a Brüel \& Kj\ae r Type 4295 omnidirectional loudspeaker and a pair of smart glasses with 8~microphones worn by a Brüel \& Kj\ae r HATS 5128-C dummy head. The microphone locations on the smart glasses are illustrated in Fig.~\ref{fig:glassesWithMicPos}. The RIRs were measured once with the dummy head looking toward the source and once with the dummy head rotated by 45$^\circ$ in azimuth resulting in 46~sets of RIRs. Additionally, all measurements were repeated with the mouth simulator of the dummy head as the source to be able to investigate the use of the user's own voice for the estimation. The rooms had volumes between \SI{70}{m^3} and \SI{1215}{m^3} (mean \SI{266}{m^3}) and the source-receiver distance (for the omnidirectional loudspeaker) varied between \SI{0.9}{m} and \SI{8.0}{m} (mean \SI{3.5}{m}). The RTs varied between \SI{0.4}{s} and \SI{1.4}{s} (mean \SI{0.6}{s}) and the DRRs between $-$\SI{12.9}{dB} and \SI{0.7}{dB} (mean $-$\SI{7.5}{dB}) as shown in Fig.~\ref{fig:dataSetRts}.

\subsection{Signal Generation and Parameter Estimation}\label{sec:objEvalSigGen}
For the evaluation of the proposed method, the measured far-field and own-voice RIRs were convolved with a source signal containing \SI{5.6}{s} of either male or female speech. To investigate the performance for different numbers of microphones, not only the full 8-channel microphone array was used but also sub-arrays using only the microphones with indices $\{1,3,5,6,8\}$ and $\{1,5,8\}$ in Fig.~\ref{fig:glassesWithMicPos}. Multichannel babble noise with the coherence of the microphone array in an isotropic diffuse field was generated using measured anechoic ATFs and the method from~\cite{Habets2008}. The noise was added to the microphone signals with varying gain to achieve average SNRs of \SI{6}{dB}, \SI{12}{dB}, and \SI{20}{dB} over all channels. The noise gain was calculated only considering signal parts where speech was active. Next, the multichannel signals were dereverberated using the GWPE method with the identity matrix approach for covariance estimation, a prediction delay of \SI{20}{ms}, a filter order of~36, and STFT block processing using a block length of 2048~samples, a square-root Hann window, and a hop size of 128~samples. A sample rate of \SI{48}{kHz} was used for all signals in this work. The MVDR beamformer coefficients were calculated using measured ATFs and the noise covariance was estimated using \SI{1}{s}
of the babble noise. The beamformer was informed of the true source DOA and applied using the same STFT processing as for the dereverberation. From the resulting pseudo reference signal and the noisy array signals, transfer functions were estimated for each microphone channel using the MWF with a regularization constant $\delta = 10^{-4}$. The STFT processing used a rectangular window, a hop size of 2048~samples, and a block length of twice the true broadband RT so that the approximation of the convolutive transfer functions as multiplicative transfer functions in the STFT domain is valid~\cite{Avargel2007a}. Note that the ground truth RT is not available in practical scenarios and the impact of deviations from the ground truth is thus investigated in Sec.~\ref{sec:robustness}. The final RIR estimates were then limited to half the block length in the time domain. This concludes the processing for the objective evaluation: the multichannel RIR estimates were directly used for the calculation of the evaluation metrics. 

The best-performing algorithms for RT~\cite{Prego2012} and DRR estimation~\cite{Hioka2015} from the ACE challenge~\cite{Eaton2016} were employed as baselines. 
The RT estimator finds free-decay regions in the signal and fits decay slopes in frequency bands to estimate frequency-dependent RTs. In contrast to the original publication where a fullband RT is obtained from the subband estimates using a mapping function, we directly use the subband estimates as final parameter estimates to be able to compare RT estimates in octave bands. The DRR estimator calculates DRRs from direct and reverberant PSDs obtained via two delay-and-sum beamformers, one pointed at the source and one rotated by $60^\circ$ in azimuth. It is provided with the true source DOA, ATFs for the beamformer design, and times of voice activity. The estimator uses all microphone signals for the estimation but only outputs a single DRR estimate. For the calculation of the DRR error, this estimate is compared to the average ground truth DRR over all channels. A bias compensation was not performed.

\subsection{Evaluation Metrics}\label{sec:object-eval-metrics}
The RT, the DRR, and the weighted angular error (WAE) were employed as evaluation metrics. 
The RTs were calculated in octave bands between \SI{125}{Hz} and \SI{8}{kHz} and the calculation of the other metrics was bandlimited between \SI{100}{Hz} and \SI{8}{kHz} where sufficient speech energy could be assumed. 

The RTs of the ground truth and the estimates were calculated for each channel of the measured and estimated RIRs using the T20-estimation method of the ITA toolbox~\cite{Berzborn2017}. The T20-estimator fits linear slopes to the sections of the energy decay curves with values between \SI{-5}{dB} and \SI{-25}{dB}. This procedure may be inaccurate if high noise levels are present and even fail if the noise floor in a certain frequency band is higher than the lower slope limit, i.e., higher than \SI{-25}{dB} for T20. To still be able to provide an estimate that can be compared to the ground truth, the result for the corresponding channel and octave band was in such cases obtained using T15 estimation, or T10 estimation if the T15 estimation also failed. If the T10 estimation was unsuccessful too, the estimation for the corresponding microphone channel and frequency band was reported as failed and not considered in the calculation of the results. The ground truth RIRs and RIR estimates from scenarios without noise do not contain high noise levels and thus the corresponding RT estimation never failed.
Note that the ground truths used for the calculation of the RT errors are not the same for all methods. While the proposed RT estimation from far-field speech and the baseline estimators use far-field signals and corresponding ground truths, the proposed method using own speech has its own ground truths, and the RT errors are calculated in relation to those.

Broadband DRRs were calculated for each channel by finding the direct sound peak and extracting the direct energy as the energy within a rectangular window starting \SI{1}{ms} before and ending \SI{2}{ms} after the peak. The reverberant energy was calculated as the energy from \SI{2}{ms} after the direct sound until the beginning of the noise floor which was determined by the RT estimator. As DRR estimates from own speech are only useful for virtual source rendering if they can be related to DRRs from far-field speech, we use the DRR ground truths from the far-field measurements even to calculate errors for the estimates from own speech.

The WAE~\cite{Meyer-Kahlen2022c} compares the directional energy distribution of the multichannel RIR estimate to the one of the ground truth. This is achieved by representing the multichannel RIRs via circular harmonic (CH) coefficients using a regularized least-squares approach (with \SI{20}{dB} maximum amplification) based on anechoically measured ATFs~\cite{Moreau2006}. Two-dimensional pseudo intensity vectors are then obtained by multiplying the omnidirectional, zeroth-degree CH component and the first-degree components, and temporally smoothing them by convolution with a 14-sample Hann window.
Finally, the angular mismatch of the pseudo intensity vectors of the estimate and the ground truth is determined and weighted by the total energy of the RIR.
As the perceptual importance of directional energy is particularly high for early reflections, the WAE was calculated for the early part of the RIRs up to \SI{50}{ms} after the direct sound. Recall that the estimation from own speech only provides RIR estimates for microphones~1 and 8. Thus, no WAE is computed for the own-speech estimates. 

As in~\cite{Eaton2016}, RT and DRR are evaluated by calculating the MSE, the bias, and the Pearson correlation coefficient $\rho$. Additionally, we investigate the median absolute error (MAE) and the median absolute deviation (MAD) as they allow for a more intuitive interpretation. The MAE of the RT is given in percent, relative to the ground truth RT.
The final error measures were obtained by averaging (MSE) or taking the median over all channels and frequency bands (where applicable).

Although these metrics correlate well with certain attributes of room acoustic perception and are commonly used for the analysis and synthesis of reverberation, they do not comprehensively capture all relevant perceptual properties like, for instance, coloration. Thus, also a listening experiment was performed (see Sec.~\ref{sec:experiment}).

\subsection{Estimation from Own Speech}\label{sec:own_voice}
Part of the proposed method has been shown to successfully estimate RIRs under noise-free conditions for speech sources in the far field and that dereverberation significantly improves the results compared to processing without dereverberation~\cite{Deppisch2024}. In this contribution, we additionally investigate the estimation of RIRs from the user's own speech, i.e., speech from the person wearing the smart glasses, as this is relevant in scenarios where no additional sound sources are available.
Thus far, it is unclear if the proposed processing is optimal for near-field sources and if near-field RIRs from the mouth are suitable for the estimation of RTs due to the expected high DRR. Thus, we investigate in this section if near-field RIRs from the user's mouth can be used to estimate RTs and if dereverberation and ATF-based near-field beamforming are beneficial for the estimation of RIRs from own speech. Results from the systematic evaluation using both own and far-field speech are presented in Sec.~\ref{sec:resultsObjective}.

None of the assumptions in the signal model and processing from Sec.~\ref{sec:rir_estimation} require far-field sources. However, for the estimation from own speech, the steering vector $\bm a(\omega)$ used in the beamformer design in~\eqref{eq:MVDR_beamformer} is replaced by an ATF that is measured using the mouth simulator of a Brüel \& Kj\ae r HATS 5128-C dummy head. Due to the static relation between the microphone array and the user's mouth, knowledge of the DOA is not required. The RIRs from the user's mouth to the array microphones only contain valuable information if the microphones are sufficiently far away from the mouth. Thus, we restrict the RIR estimation to the microphones closest to the user's ears (microphones~1 and 8 in Fig.~\ref{fig:glassesWithMicPos}) in all investigations that utilize the user's own speech. 

To quantify if measured near-field RIRs can be used to estimate RTs and if these estimates correlate with RTs from far-field measurements, we compared RTs from far-field RIR measurements in the 23~rooms to RTs from RIR measurements using the mouth simulator of the dummy head wearing the smart glasses. We found a median absolute difference between the RTs from mouth and far-field measurements of \SI{11}{\%} of the far-field RT, a correlation coefficient of $0.91$, and a bias of \SI{-0.06}{s}, showing that the use of own speech for RT estimation warrants further exploration.

\begin{figure}[t]
  \centering
  \includegraphics[width=\columnwidth]{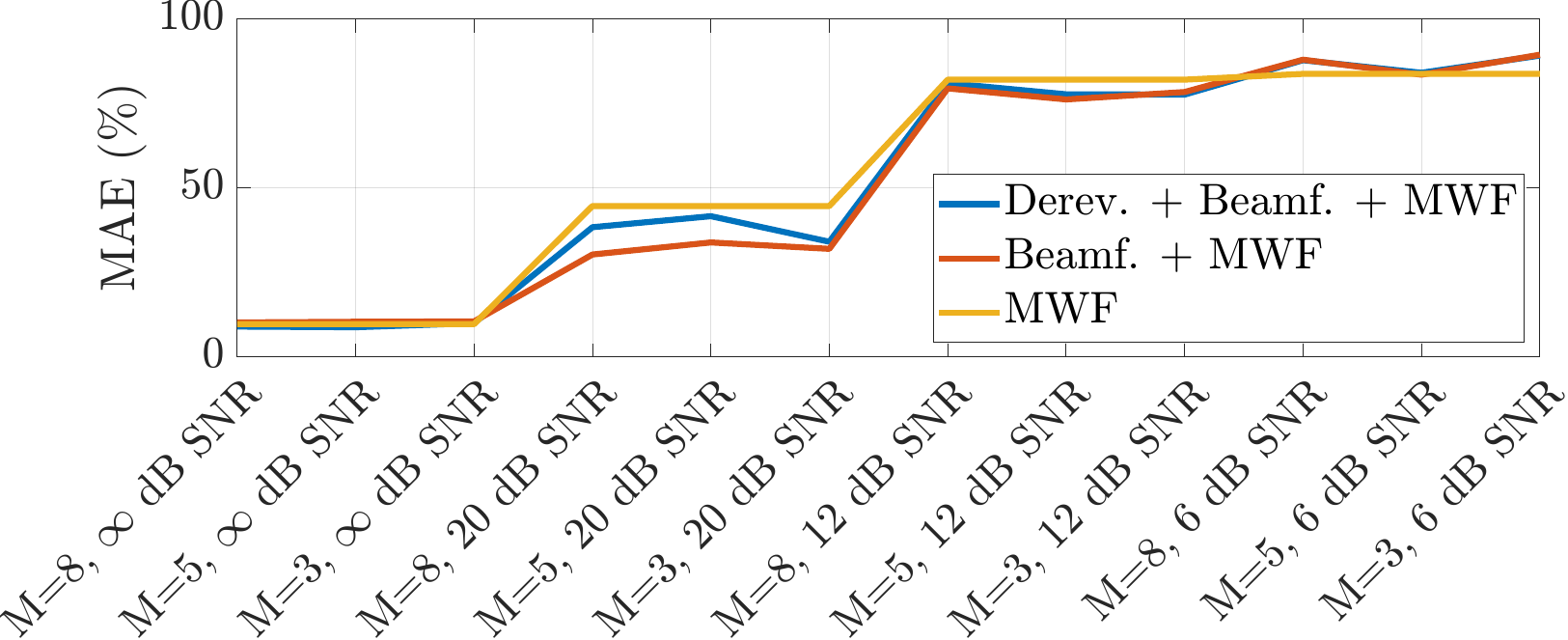}
  \caption{MAEs of RTs for RIR estimates obtained from the user's own speech for varying numbers of microphones $M$ and SNRs.}
  \label{fig:ownVoicePreStudy}
\end{figure}

To determine the most suitable processing chain, we compare RIR estimates from the full chain to estimates obtained without dereverberation and to estimates obtained without dereverberation and beamforming. In the latter case, the signal from the microphone closest to the user's mouth (microphone~5 in Fig.~\ref{fig:glassesWithMicPos}) is utilized as the pseudo reference signal.
Fig.~\ref{fig:ownVoicePreStudy} shows MAEs of the RT estimates of the three configurations of the method for varying numbers of microphones $M$ and SNRs. Only at an SNR of \SI{20}{dB}, a significant difference between the different configurations is observed. Here, the configuration using beamforming but no dereverberation performs best with median errors between \SI{30}{\%} and \SI{34}{\%}. As some of the microphones are located close to the user's mouth, a sufficiently clean reference is captured without the additional dereverberation. This configuration is thus used to obtain the results in Sec.~\ref{sec:resultsObjective}. 

Without noise, all three methods perform similarly, resulting in MAEs of around \SI{10}{\%}. At SNRs of \SI{12}{dB} and lower, all methods create large errors of about \SI{80}{\%}. Interestingly, even in the scenarios with a rather high SNR of \SI{20}{dB}, the additional noise considerably impairs the results compared to the scenarios without noise. This likely happens because of the proximity of the mouth to the microphones, resulting in much lower reverberant signal energy, which is used to estimate RTs, than direct signal energy. Note however that high-SNR scenarios are most relevant for the estimation from own speech as the proximity of the user's mouth to the microphones results in a high SNR even if the user is located in a noisy environment. To quantify this effect, we conducted measurements under anechoic conditions comparing the speech level when the speaker wears the smart glasses to the same speech in \SI{2}{m} distance and found about \SI{20}{dB} higher sound pressure levels for frequencies up to \SI{1}{kHz} in the first case and about \SI{10}{dB} up to \SI{6}{kHz}. Thus, an SNR of \SI{20}{dB} for the estimation from own speech is more comparable to an SNR of \SI{0}{dB} for far-field speech.

\subsection{Results}\label{sec:resultsObjective}
The results from the proposed method using far-field and own speech and from the baseline methods are presented in Tab.~\ref{tab:evalResults} for different array configurations and SNRs
including results for a best-case scenario without noise and $M=8$ microphones.

\begin{table*}[t]
\centering
\caption{Evaluation results of the proposed method (MWF) using far-field speech, using own speech, and the best-performing algorithms from the ACE challenge~\cite{Prego2012,Hioka2015} (where applicable) with varying numbers of microphones $M$ and SNRs.}
\resizebox{\textwidth}{!}{%
\begin{tabular}{ll|llllll|lllll|ll}
\hline
 &  & \multicolumn{5}{l}{\textbf{RT}} &  & \multicolumn{5}{l|}{\textbf{DRR}} & \multicolumn{2}{l}{\textbf{WAE}} \\
 &  & MSE $\downarrow$ & bias $\lvert\downarrow\rvert$ & $\rho$ $\uparrow$ & MAE (\%) $\downarrow$ & MAD (\%) $\downarrow$ & failed (\%) & MSE $\downarrow$ & bias $\lvert\downarrow\rvert$ & $\rho$ $\uparrow$ & MAE (dB) $\downarrow$ & MAD (dB) $\downarrow$ & MAE ($^\circ$) $\downarrow$ & MAD ($^\circ$) $\downarrow$ \\ \hline
  & MWF Far Speech & 0.03 & \textbf{0.01} & 0.89 & \textbf{8.85} & 5.22 & 0 & \textbf{5.11} & \textbf{0.21} & \textbf{0.83} & \textbf{1.29} & \textbf{0.84} & 44.36 & 7.95 \\
M=8, $\infty$ dB SNR & MWF Own Speech & \textbf{0.01} & 0.07 & \textbf{0.95} & 9.98 & \textbf{4.24} & 0 & 248.49 & -15.28 & 0.4 & 14.82 & 2.31 &  &  \\
 & ACE \cite{Prego2012, Hioka2015} & 0.08 & -0.09 & 0.68 & 16.19 & 10.63 & 0 & 89.33 & 9.19 & 0.62 & 8.94 & 1.35 &  &  \\ \hline
 & MWF Far Speech & \textbf{0.03} & \textbf{0.03} & \textbf{0.88} & \textbf{10.22} & \textbf{6.15} & 0.02 & \textbf{4.6} & \textbf{0.3} & \textbf{0.83} & \textbf{1.17} & \textbf{0.78} & 45.46 & 9.93 \\
M=8, 20 dB SNR & MWF Own Speech & 0.11 & 0.16 & 0.72 & 28.65 & 23.79 & 7.22 & 94.88 & -9.15 & 0.45 & 9.07 & 2.04 &  & \\
 & ACE \cite{Prego2012, Hioka2015} & 0.12 & -0.2 & 0.62 & 29.42 & 21.6 & 0 & 89.55 & 9.2 & 0.62 & 8.96 & 1.37 &  & \\ \hline
 & MWF Far Speech &\textbf{0.03} & \textbf{0.05} & \textbf{0.85} & \textbf{11.35} & \textbf{7.25} & 1.32 & \textbf{4.65} & \textbf{0.3} & \textbf{0.83} & \textbf{1.18} & \textbf{0.79} & 46.12 & 9.6 \\
M=8, 12 dB SNR & MWF Own Speech & 0.25 & 0.39 & 0.57 & 78.46 & 14.71 & 31.37 & 66.51 & -7.38 & 0.4 & 6.98 & 2.55 &  &  \\
 & ACE \cite{Prego2012, Hioka2015} & 0.19 & -0.26 & 0.47 & 35.71 & 26.87 & 0 & 90.43 & 9.24 & 0.62 & 9.03 & 1.32 &  &  \\ \hline
 & MWF Far Speech & \textbf{0.04} & \textbf{0.08} & \textbf{0.82} & \textbf{13.4} & 8.67 & 6.29 & \textbf{4.66} & \textbf{0.33} & \textbf{0.83} & \textbf{1.2} & \textbf{0.79} & 45.67 & 9.5 \\
M=8, 6 dB SNR & MWF Own Speech & 0.43 & 0.55 & 0.49 & 87.8 & \textbf{8.56} & 66.54 & 29.55 & -4.07 & 0.42 & 4.46 & 2.33 &  &   \\
 & ACE \cite{Prego2012, Hioka2015} & 0.25 & -0.31 & 0.35 & 48.76 & 36.19 & 0 & 91.55 & 9.28 & 0.58 & 9.07 & 1.38 &  &  \\ \hline
 & MWF Far Speech & \textbf{0.04} & \textbf{0.02} & \textbf{0.85} & \textbf{11.16} & \textbf{6.94} & 0.12 & \textbf{6.34} & \textbf{-0.32} & \textbf{0.76} & \textbf{1.62} & \textbf{0.93} & 35.25 & 8.11 \\
M=5, 20 dB SNR & MWF Own Speech & 0.12 & 0.17 & 0.74 & 32.75 & 25.26 & 8.15 & 75.83 & -7.98 & 0.38 & 7.67 & 1.91 &  & \\
 & ACE \cite{Prego2012, Hioka2015} & 0.12 & -0.2 & 0.62 & 28.77 & 21.28 & 0 & 92.7 & 9.38 & 0.66 & 9.15 & 1.56 &  & 
 \\ \hline
 & MWF Far Speech & \textbf{0.05} & \textbf{0.04} & \textbf{0.82} & \textbf{12.45} & \textbf{7.98} & 1.49 & \textbf{6.45} & \textbf{-0.34} & \textbf{0.76} & \textbf{1.61} & \textbf{0.94} & 34.42 & 8 \\
M=5, 12 dB SNR & MWF Own Speech & 0.23 & 0.38 & 0.62 & 74.19 & 17 & 31.99 & 50.88 & -6.38 & 0.52 & 6.26 & 1.8 &  & \\
 & ACE \cite{Prego2012, Hioka2015} & 0.2 & -0.26 & 0.47 & 35.57 & 26.47 & 0 & 97.29 & 9.61 & 0.65 & 9.39 & 1.57 &  & \\ \hline
 & MWF Far Speech & \textbf{0.06} & \textbf{0.08} & \textbf{0.79} & \textbf{14.92} & 9.55 & 6.3 & \textbf{6.44} & \textbf{-0.31} & \textbf{0.76} & \textbf{1.62} & \textbf{0.96} & 34.83 & 7.36 \\
M=5, 6 dB SNR & MWF Own Speech & 0.35 & 0.5 & 0.56 & 82.45 & \textbf{9.22} & 63.66 & 23.38 & -3.66 & 0.55 & 3.63 & 1.99 &  & \\
 & ACE \cite{Prego2012, Hioka2015} & 0.26 & -0.32 & 0.34 & 50.33 & 37.64 & 0 & 111.86 & 10.33 & 0.61 & 10 & 1.52 &  & \\ \hline
 & MWF Far Speech & \textbf{0.11} & \textbf{-0.08} & \textbf{0.87} & \textbf{17.19} & \textbf{11.62} & 0.16 & \textbf{16.7} & \textbf{-2.14} & 0.58 & \textbf{2.76} & \textbf{1.43} & 32.89 & 7.69 \\
M=3, 20 dB SNR & MWF Own Speech & 0.13 & 0.18 & 0.7 & 33.48 & 28.14 & 10.02 & 217.55 & -14.29 & 0.39 & 14.04 & 1.92 &  & \\
 & ACE \cite{Prego2012, Hioka2015} & 0.12 & -0.19 & 0.61 & 27.8 & 21.04 & 0 & 72.9 & 8.24 & \textbf{0.68} & 7.99 & 1.47 &  &  \\ \hline
 & MWF Far Speech & \textbf{0.09} & \textbf{-0.02} & \textbf{0.83} & \textbf{18.57} & \textbf{12.5} & 1.6 & \textbf{18.45} & \textbf{-2.39} & 0.57 & \textbf{3.06} & 1.66 & 32.75 & 8.04 \\
M=3, 12 dB SNR & MWF Own Speech & 0.27 & 0.39 & 0.56 & 79.28 & 16.89 & 38.35 & 141.86 & -11.44 & 0.49 & 11.53 & 2.25 &  & \\
 & ACE \cite{Prego2012, Hioka2015} &0.19 & -0.26 & 0.47 & 36.95 & 27.68 & 0 & 76.37 & 8.45 & \textbf{0.68} & 8.17 & \textbf{1.56} &  & \\ \hline
 & MWF Far Speech & \textbf{0.08} & \textbf{0.04} & \textbf{0.79} & \textbf{20.17} & 12.6 & 8.28 & \textbf{18.71} & \textbf{-2.45} & 0.56 & \textbf{3.2} & 1.72 & 32.73 & 8.42\\
M=3, 6 dB SNR & MWF Own Speech & 0.43 & 0.55 & 0.55 & 88.34 & \textbf{9.58} & 72.28 & 67.61 & -7.5 & 0.48 & 7.4 & 2.5 &  & \\
 & ACE \cite{Prego2012, Hioka2015} & 0.26 & -0.33 & 0.34 & 51.88 & 39.64 & 0 & 88.13 & 9.11 & \textbf{0.67} & 8.78 & \textbf{1.57} &  & \\ \hline
\end{tabular}%
}
\label{tab:evalResults}
\end{table*}

\paragraph{Reverberation Time (RT)}
The proposed method outperforms the baseline estimator in all scenarios and RT metrics. The method achieves the best results in terms of MAE and bias in all test cases when using far-field speech. It achieves the highest correlation coefficients in most cases, only in the noise-free scenario the proposed method using own speech generates an even higher correlation. The generated MAEs are with \SI{8.85}{\%} lowest in the scenario without noise and $M=8$ microphones. They increase with decreasing SNR and a decreasing number of microphones to a maximum MAE of \SI{20.17}{\%} for $M=3$ and an SNR of \SI{6}{dB}. When the method uses the wearer's own speech, it achieves good results in the high-SNR scenarios. Without the presence of noise, it creates an MAE of \SI{9.98}{\%}, clearly outperforming the baseline algorithm with an MAE of \SI{16.19}{\%}. At lower SNRs, the method using own speech creates significantly higher errors than the method using the far-field speech but at an SNR of \SI{20}{dB}, its results are comparable to the ones from the baseline algorithm from the ACE challenge creating MAEs around \SI{30}{\%}. At lower SNRs, the MAEs of both the own speech and the baseline estimator increase strongly but the ACE estimator creates significantly lower errors than the method using own speech. However, the method using own speech in all cases shows a higher correlation coefficient so its results may be improved by a bias compensation.

\begin{figure}[t]
  \centering
  \includegraphics[width=\columnwidth]{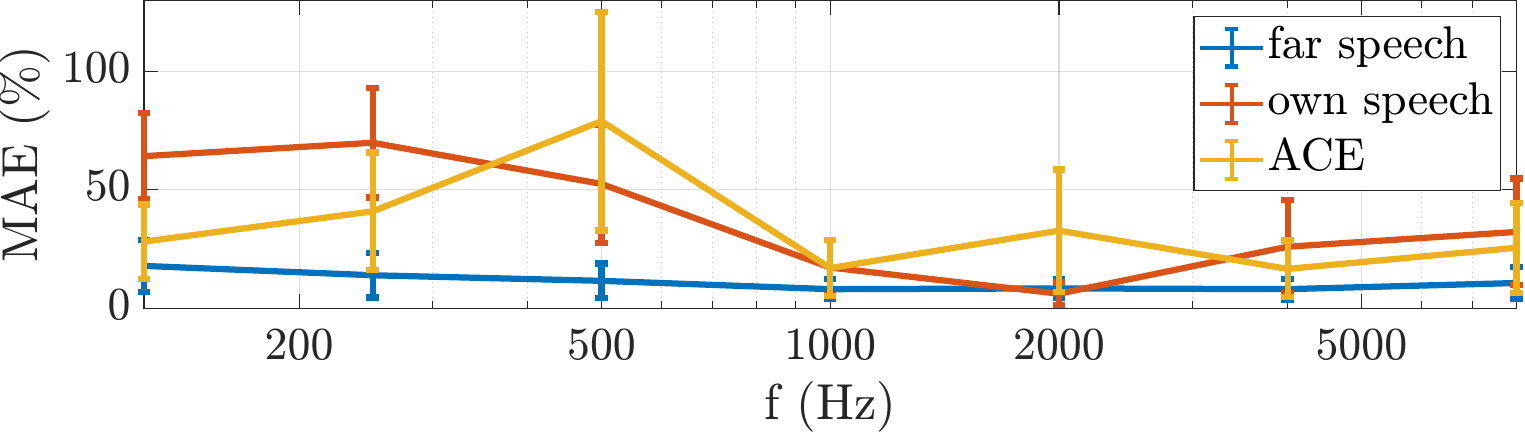}
  \caption{Frequency-dependent median absolute reverberation time errors for estimates obtained from the proposed method and far-field speech, own speech, and the ACE estimator~\cite{Prego2012}.}
  \label{fig:medianRtErrorFreqDep}
\end{figure}

Fig.~\ref{fig:medianRtErrorFreqDep} shows the MAEs as a function of frequency for $M=8$ microphones and an SNR of \SI{20}{dB}. While the proposed method using far-field speech generates low errors across the full frequency range, with only slightly increasing MAEs toward low frequencies, the results from the other methods fluctuate more strongly. The MAEs from the proposed method using own speech strongly increase toward low frequencies and the baseline estimator creates substantial errors at the \SI{500}{Hz} and \SI{2}{kHz} octave bands.

As described in Sec.~\ref{sec:object-eval-metrics}, cases may exist where the RT estimation fails and hence the last column of the RT metrics in Tab.~\ref{tab:evalResults} shows the percentage of such cases. With the proposed method using far-field speech, \SI{0.02}{\%} of estimates failed in the scenario with \SI{20}{dB} SNR $M=8$ microphones. This percentage increases with lower SNRs and less microphones, to a maximum of \SI{8.28}{\%} at an SNR of \SI{6}{dB} and $M=3$ microphones. The RT estimation failed in a higher proportion of cases when using own speech, ranging between \SI{7.22}{\%} and \SI{72.28}{\%}. As discussed in Sec.~\ref{sec:own_voice}, this likely is caused by the lower ratio of reverberant to direct energy. In practice, a low amount of failed RT estimates is typically tolerable as it can be compensated for by interpolation from neighboring frequency bands or microphone channels. Failure to estimate the RT with the traditional slope fitting may also be avoided by employing other methods that are more robust against a high noise floor such as~\cite{Gotz2022a}. The ACE estimator uses a different fitting procedure and thus none of its estimations failed.

\paragraph{Direct-to-Reverberant Energy Ratio (DRR)}
The proposed method with far-field speech also achieves the best results for DRR estimation in almost all cases. Only in the scenarios with $M=3$ microphones, the DRR estimator from the ACE challenge reaches a higher correlation coefficient and in two of those cases also a lower MAD. In those cases, the estimator may achieve better results than the proposed method in terms of the MAE if a data-driven bias compensation is applied. However, the proposed method with far-field speech generates the lowest MAEs in all scenarios. Its MAEs change noticeably with the number of microphones but only slightly with SNR, achieving the lowest MAEs of about \SI{1.2}{dB} for $M=8$, about \SI{1.6}{dB} for $M=5$, and the highest MAEs of about \SI{3}{dB} for $M=3$.

\begin{figure}[t]
  \centering
  \includegraphics[width=\columnwidth]{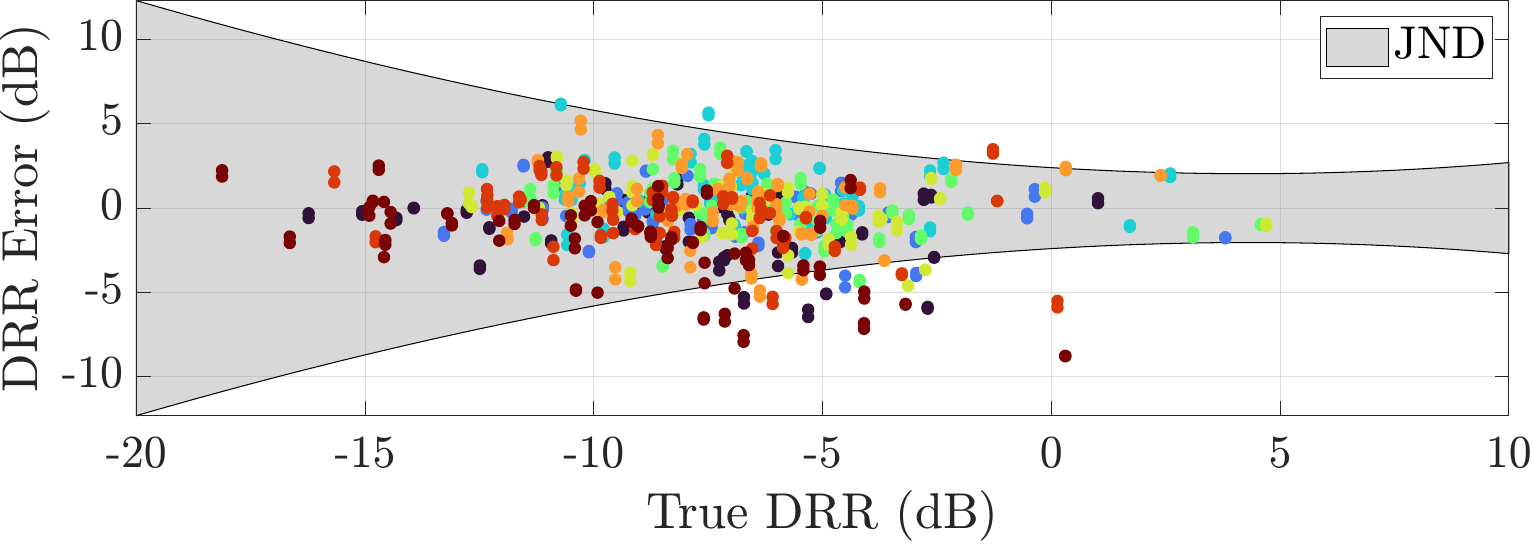}
  \caption{DRR errors obtained from the proposed method with far-field speech, $M=8$ microphones and an SNR of \SI{20}{dB}. \SI{93.2}{\%} of the DRR errors are within the JND. Colors illustrate microphone channels.}
  \label{fig:drrScatterPlot}
\end{figure}

The DRR estimates from the proposed method using far-field speech with $M=8$ microphones and an SNR of \SI{20}{dB} are shown for all microphone channels and frequency bands as a function of the true DRR in Fig.~\ref{fig:drrScatterPlot}. By comparing it to the just noticeable differences (JNDs) from~\cite{Larsen2008}, we find that \SI{93.2}{\%} of the DRR estimates are within the JND.

As expected, the results from the own speech estimation are strongly biased as the DRR from near-field speech is generally higher than the one from far-field speech. Due to the low correlation coefficient, a simple bias compensation is not sufficient to improve these results but a more sophisticated bias compensation that exploits a source distance estimate may be successful. 

\paragraph{Weighted Angular Error (WAE)}
WAEs were only computed for the proposed method using far-field speech, as the RIR estimates from own speech only comprise two microphone channels and no corresponding baseline estimator exists for further comparison. The WAEs can, however, be compared to the ones computed in~\cite{Deppisch2024}, where the proposed method was used in noise-free scenarios with different arrays of regularly distributed microphones. The most similar array from that study is an equatorial microphone array with 6~microphones, achieving a mean WAE of about $33^\circ$ which is comparable to the median WAE of about $35^\circ$ that is obtained in the present study with the smart glasses and 5~microphones. The results do not strongly depend on the SNR but rather on the number of microphones and MAEs in scenarios with more microphones tend to be higher than in scenarios with fewer microphones. Note that this does not imply a higher directional accuracy when fewer microphones are employed but only a better alignment with the ground truth which was obtained from the same set of microphones as the estimate. The expected WAE generated by a pseudo intensity vector pointing in uniformly distributed random directions is $90^\circ$.


\subsection{Robustness}\label{sec:robustness}
\begin{figure}[t]
  \centering
  \includegraphics[width=\columnwidth]{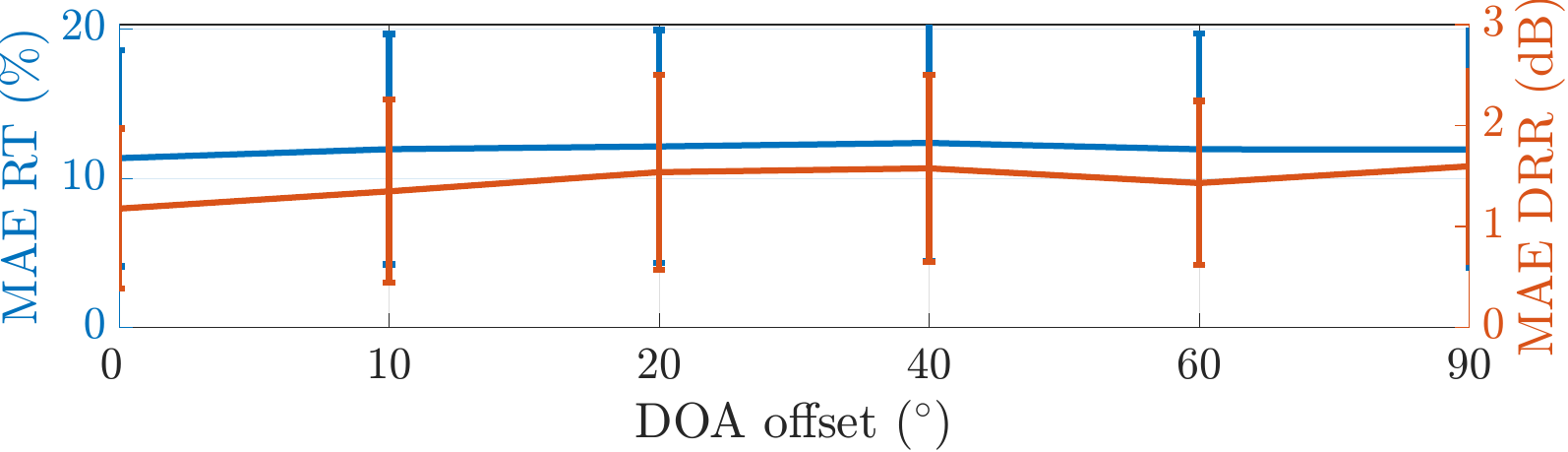}\\

  \includegraphics[width=\columnwidth]{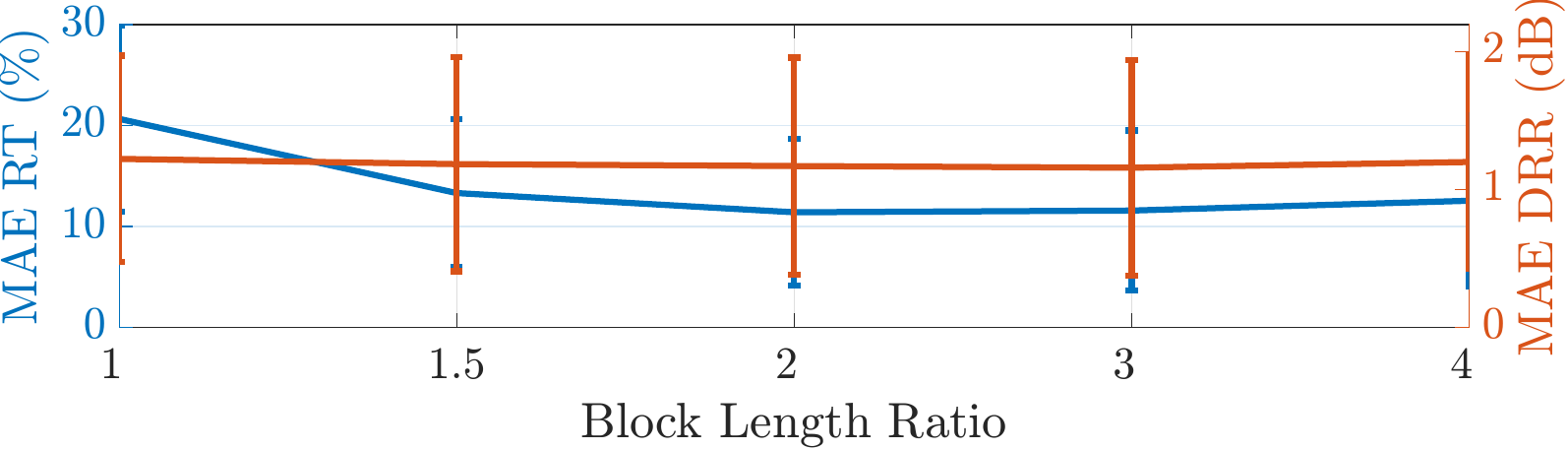}\\

  \includegraphics[width=\columnwidth]{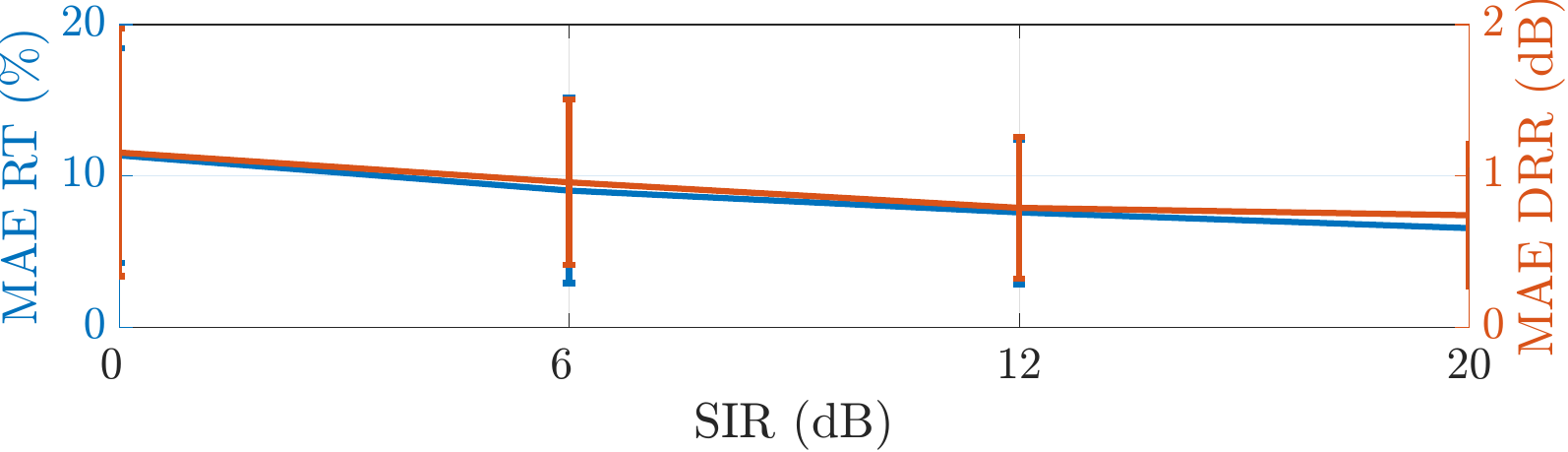}
\caption{Influence of DOA offset (top), MWF block length (center), and interfering speech (bottom), on the MAEs of RT (blue, left axis) and DRR (red, right axis) when using far-field speech.}\label{fig:robustness}
\end{figure}

We investigate the robustness of the proposed method with far-field speech against a DOA offset, deviations in the block length of the MWF, and interfering speech. The influence on MAEs of RT and DRR for $M=8$ microphones are shown in Fig.~\ref{fig:robustness}. 

While in the previous analysis, the MVDR beamformer used an ATF from the true source direction as a steering vector, here an ATF from a direction with an azimuthal offset of varying angle to the true source direction was used. The results were obtained for an SNR of \SI{12}{dB}. With an increased DOA offset, the MAEs increase slightly from about \SI{11}{\%} to \SI{12}{\%} for the RT estimates and more significantly from \SI{1.2}{dB} to \SI{1.6}{dB} for DRRs, indicating that RTs can be estimated even if the direct sound is not captured accurately. In contrast, the DRR estimates benefit more from an accurate DOA estimate.

The choice of the block length of the MWF has a larger influence on the MAEs of RTs. On the other hand, it only has a minor influence on MAEs of DRRs. In the previous evaluation, the MWF block length was chosen to be equal to twice the ground truth RT which is shown to give the best results in Fig.~\ref{fig:robustness}~(center). When choosing the length to be in the range between $1.5$ to $4$ times the ground truth RT, the MAEs of RTs only increase slightly by up to \SI{2}{\%}. When the MWF block length is equal to the ground truth RT, the MAE increases by about \SI{9}{\%}. 

Lastly, we consider a scenario with interfering speech but no additional diffuse noise. For this investigation, a set of 3~rooms was considered, where measurements for source directions at $-30^\circ$, $0^\circ$, and $30^\circ$ were available. Measurements from the same rooms were used for the perceptual evaluation and are discussed in more detail in Sec.~\ref{sec:experimentDataSet}. The results were obtained for all combinations of positions as target speech position and interfering speech position, resulting in 6~combinations per room and 18~total combinations. The estimation was performed with male target speech and female interfering speech and vice versa. Fig.~\ref{fig:robustness} (bottom) shows MAEs for varying signal-to-interference ratios (SIRs), i.e., varying energy of the interfering speech compared to the speech that is considered for the estimation. For example, at an SIR of \SI{0}{dB}, target speech and interfering speech have equal energy. Note that we assume no knowledge about the interferer, neither in terms of its DOA nor its PSD. The MAEs of both RT and DRR moderately increase with decreasing SIR, from \SI{7}{\%} to \SI{12}{\%} and \SI{0.7}{dB} to \SI{1.2}{dB}, respectively.




\section{Listening Experiment}\label{sec:experiment}
Apart from the objective evaluation, the resynthesized BRIRs should be evaluated perceptually. 
With the AR application in mind, real/virtual tests under the plausibility~\cite{Lindau2012} or transfer-plausibility paradigm~\cite{Wirler2020}, comparing a headphone-rendered stimulus to a stimulus that is played back with a loudspeaker in the room, may be considered appropriate. However, performing real/virtual tests using multiple rooms is cumbersome as participants need to be individually tested in all rooms and the rooms need to be equipped according to the experimental demands. Moreover, the experiment should investigate the perceptual quality of the generated BRIRs without the impact of other factors like non-individualized HRTFs, head-tracking artifacts, and the distortion of external sounds through headphones. Thus, we only use static, headphone-rendered stimuli intending to answer two research questions:

(i) Is it better to use an estimated response than a measured response from another room of similar size?

(ii) Is the estimated response as similar to a measured response as a measured response from another position in the same room?

Therein, (i) aims to test whether BRIR estimation is worth the effort when the alternative is to choose an available response from a different, but similar room for rendering. Question (ii) aims to assess if the quality of the estimate is as good as a measurement.

\subsection{Experiment Design}
To answer questions (i) and (ii), a test was designed with a reference sample and two test samples in each trial. 
Participants were asked to listen to the three samples and to select which one of the test samples (A or B) sounded as if it was \emph{recorded in the same room} as the reference. The reference used one speech signal as source material and the test samples used a different speech signal. Different speech signals for different samples in one trial are also employed in real/virtual tests under the transfer-plausibility paradigm~\cite{Wirler2020}. Following~\cite{Mckenzie2023}, where, like here, all renderings were virtual, the measurement from the same position rendered with a different signal would be called a \emph{transfer-reference}. Also, the design is akin to the identification test performed within the co-immersion framework \cite{fantini_co-immersion_2023}.

\begin{figure}[t]
  \centering
  \includegraphics[width=0.9\columnwidth]{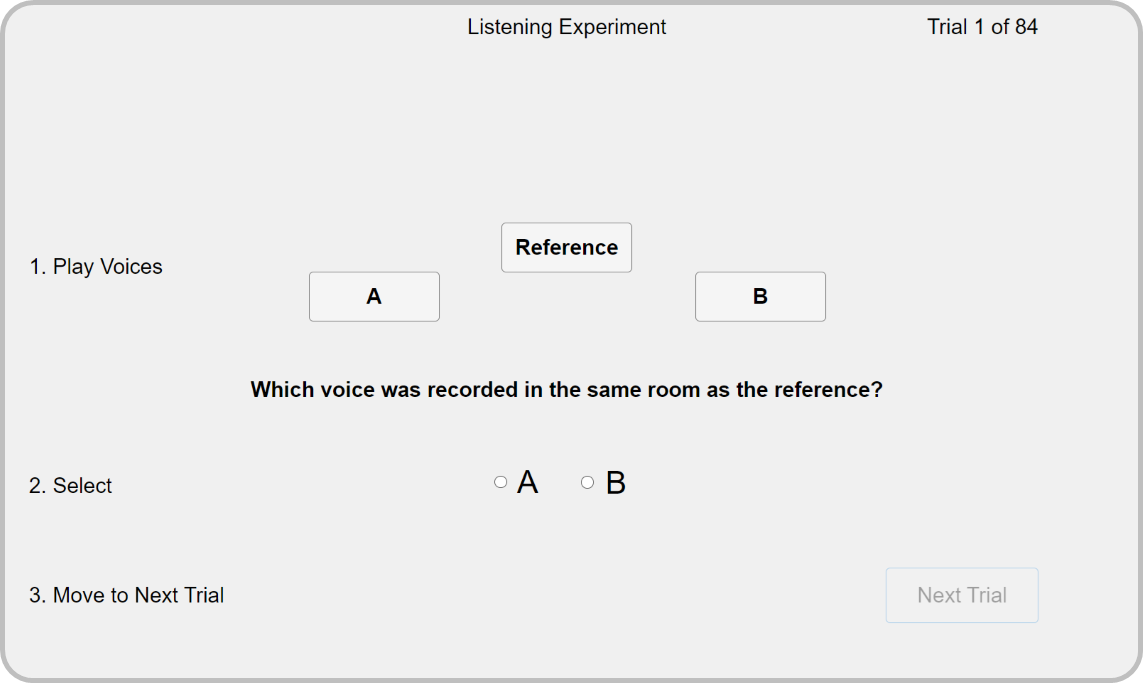}
  \caption{User interface of the listening experiment.}
  \label{fig:expInterface}
\end{figure}

The user interface of the experiment is shown in Fig.~\ref{fig:expInterface}. In all trials of all parts of the experiment, the speech signals of the reference and the test samples were different but of the same type (female or male speech) and were based on BRIRs from different positions. For example, if the reference sample was created from BRIRs with the source located $30^\circ$ to the left of the listener convolved with a sample of female speech, the test samples would be generated using either measured BRIRs or estimated BRIRs with the source located $30^\circ$ to the right of the listener, convolved with the same audio sample of the female speaker but this speech sample would differ from the one used by the reference. 

To answer~(i), two parts were included. In part I.a, the reference was based on a measured response, and the first test stimulus was based on an estimated response from the same room as the reference, whereas the second test stimulus was based on a measured response from a different room. Thus, choosing the estimated response implies that it is closer to the reference than the response of the other room. If participants choose the measured response from another room, it remains unclear if the confusion originated from shortcomings of the estimate or from the fact that the tested rooms are so similar that participants cannot distinguish them. Therefore, part I.b tests if participants consistently select a response based on a measurement from the same room over a measurement from a different room.

To assess the much stricter question (ii), part II of the experiment used a measured response from one position in a room for generating the reference. One of the test samples was then based on a measured response from a different location in the same room and the other one on an estimated response from the room. Here, participants are expected to only confuse the measurement with the estimate if the estimate reproduces RT and DRR with low error and is completely free from artifacts and audible degradations.

The experiment comprised a total of 84~trials. Parts I.a and I.b of the experiment were initially designed with 24~trials each, containing the combinations of 6~room comparisons (each of the 3~rooms compared to 2 other rooms), 2~source positions, and 2~signal types (female or male speech). After a pilot test, a repetition of all trials of part I.a with different speech signals was added to reduce the influence of individual speech signals, resulting in 72~trials for part I of the experiment. Part II of the experiment only contained within-room comparisons of measured and estimated responses and thus had 12~trials including all combinations of the 3~rooms, 2~positions, and 2~signal types. All trials appeared in random order for each participant and the trials of parts I and II were not separated. Within each trial, test samples A and B were randomly assigned to buttons A and B in the user interface for each participant. To familiarize the participants with the procedure, they had to complete 3~training trials whose answers were not included in the results.

\subsection{Data Set}\label{sec:experimentDataSet}
For the listening experiment, a data set with RIRs from geometrically similar rooms with the same relative source-receiver positions was required. The chosen data set comprises three sets of RIRs, obtained from a meeting room of dimensions $6.0 \times 5.9 \times 2.7$~\si{m} and a variable-acoustics lab space of dimensions $9.7 \times 5.5 \times 2.7$~\si{m} that was measured in a \emph{dry} condition, i.e., with acoustic wall panels turned to their absorbing side, and in a \emph{reverberant} condition with the panels turned to their reflecting side. Within each room, two source positions were measured with Genelec 8320A loudspeakers that were located at $\pm 30^\circ$ azimuth and a distance of \SI{3.3}{m} (meeting room) or \SI{2.7}{m} (lab space) from the Brüel \& Kj\ae r HATS 5128-C dummy head wearing the smart glasses as receiver. In the meeting room, a table was located between the loudspeakers and the dummy head. To be able to answer (i), rooms with similar dimensions were chosen so that, based on the geometry, it would be reasonable to employ a response from one of the rooms to auralize any of the others.

\begin{figure}[t]
  \centering
  \includegraphics[width=\columnwidth]{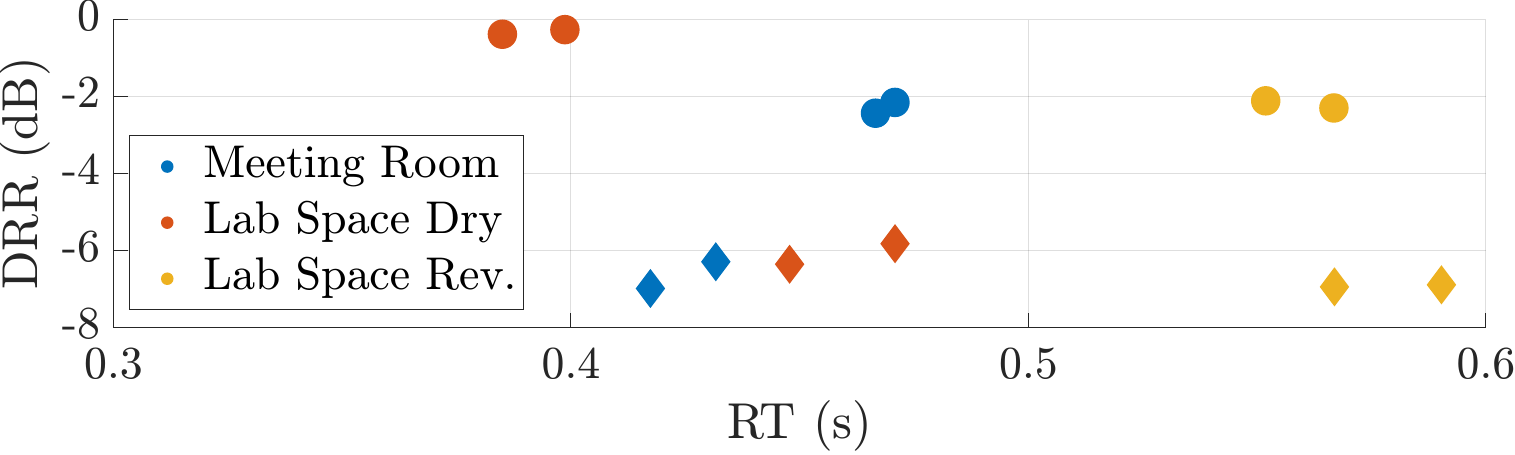}
  \caption{Average RTs and DRRs of the rooms used in the listening experiment, band-limited between \SI{100}{Hz} and \SI{8}{kHz} (circles) and between \SI{100}{Hz} and \SI{1}{kHz} (diamonds), where speech has most of its energy.}
  \label{fig:expDataSetRts}
\end{figure}

Fig.~\ref{fig:expDataSetRts} shows measured RTs and DRRs of the three rooms, averaged over microphone channels and octave bands. Circles illustrate the metrics for a frequency range between \SI{100}{Hz} and \SI{8}{kHz}, and diamonds for a limited frequency range between \SI{100}{Hz} and \SI{1}{kHz} where speech carries most of its energy. When considering the frequency range up to \SI{8}{kHz}, the three rooms have different RTs of \SI{0.39}{s} (Lab Space Dry), \SI{0.47}{s} (Meeting Room) and \SI{0.55}{s} (Lab Space Rev.) but for the limited frequency range up to \SI{1}{kHz}, Meeting Room and Lab Space Dry have similar RTs and all three rooms have similar DRRs.

\subsection{Signal Generation}
The reference samples for the experiment were generated by rendering a measured multichannel RIR using the eMagLS binaural rendering technique (cf. Sec.~\ref{sec:binauralRendering}) and convolving the resulting BRIR with a speech sample. For each trial, one out of eight speech samples of either male or female speech was picked randomly for the reference and one of the seven remaining samples of the same type (male or female) was randomly chosen for the test samples. The selection of speech samples was done beforehand so that the samples in each trial were the same for all participants. 

The multichannel RIR estimates were generated using the same parameters as for the objective evaluation in Sec.~\ref{sec:objEvalSigGen}. However, only estimates from far-field speech using the full array of 8~microphones were used and no noise was added to the array signals. The RIR estimates were then resynthesized and rendered as detailed in Sec.~\ref{sec:rendering}. All samples were band-limited to frequencies between \SI{100}{Hz} and \SI{16}{kHz}. The binaural rendering was done for forward-facing head orientation and played back via Beyerdynamic DT 770 Pro headphones without head tracking.

\subsection{Results}
23~participants aged between 24 and 62~years (average 37 years) took part in the experiment. Six~of the participants had previously participated in a listening experiment. One participant reported mild hearing loss. They took between 12 and 54 minutes (average 24 minutes) to complete the experiment.
We preferred mostly inexperienced listeners for this test as we believe that their perception will matter more in an application scenario. Also, through the test design as a two-alternative forced choice (2AFC) task with a simple question, no extensive acoustical expertise was required.

\begin{figure}[t]
\begin{subfigure}[t!]{0.45\textwidth}
\centering
\includegraphics[height=6cm, trim = {0, 0, 0, 0.7cm}, clip]{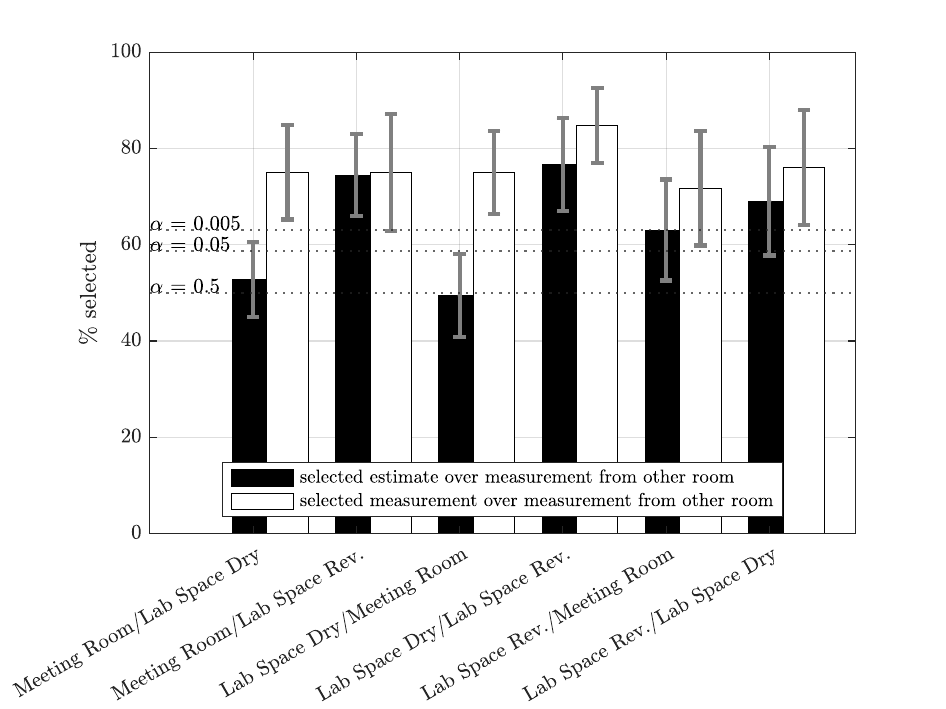}\caption{Part I - Percentages of renderings based on estimates (black) or measurements (white) selected over renderings from measurements from a different room. }
\label{fig:partIab}
\end{subfigure}
\begin{subfigure}[t!]{0.45\textwidth}
\centering
\includegraphics[height=6cm, trim = {0cm, 0, 0, 0.5cm}, clip]{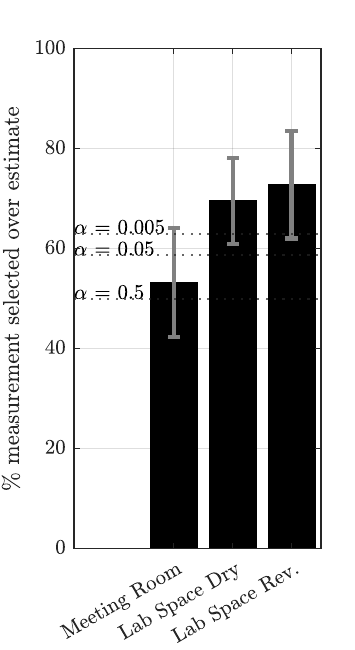}
\caption{Part II - Percentages of renderings based on measurements selected over renderings from estimates from the same room.}
\label{fig:partII}
\end{subfigure}
\caption{Results of the listening experiment. Error bars show binomial confidence intervals. Dashed lines indicate different levels of confidence $\alpha$ for the choice not originating from chance.}
\end{figure}

Fig.~\ref{fig:partIab} shows the results for part~I of the experiment. Recall that participants were asked which of the test samples \emph{was recorded in the same room} as the reference. The bars show the percentage of times an estimated response (black bars) or a measurement from the same room (white bars) was selected over a measurement from another room. The confidence intervals are indicated by error bars and computed using the Clopper-Pearson method. In addition, significance levels are indicated by the horizontal, dotted lines. These limits were computed from the binomial cumulative density function, given the overall number of trials presented in the respective condition. A confidence level of $\alpha = 0.05$ indicates that with a probability of $1-\alpha = 0.95$, the results were not obtained from randomly selecting one of the two options with equal probability.

The black bars show that for four out of six pairs, the rendering using the estimate was selected significantly more often ($p \leq 0.005$) than the rendering using a measured response from another room.
For the two pairs of responses from the Meeting Room and the dry Lab Space, the percentage of selections was close to \SI{50}{\%} which is the percentage that would result from random guessing.

The white bars show that for all pairs, a measurement from the same room was selected significantly more often ($p \leq 0.005$) than a measurement from another room. This indicates that, in principle, participants were better than chance when distinguishing between the rooms, independent of the pair. Note however that the proportions are in a range of $71$--\SI{85}{\%}, indicating that the participants did not always recognize the measurement from the same room.



Fig.~\ref{fig:partII} shows the results of part~II of the experiment, where measurements and estimates from the same position in the same room were directly compared to the reference, which was generated from a measurement from another position in the same room. For the Meeting Room, the estimates were selected about as often as the measurements. For the Lab Space in the dry and reverberant settings, the percentages of selecting the measurement over the estimate are \SI{70}{\%} and \SI{73}{\%}, which is significantly more often than it would be obtained by chance ($p \leq 0.005$).

\subsection{Discussion}
Regarding question (i), whether it is better to use an estimated response than a measured response from another room of similar size, the result is positive for four out of six pairs of rooms. Here, the estimate is more similar to the reference than a measurement from another room. For the pair Meeting Room / Lab Space Dry, using the estimate was as good as using a measurement from the other room. This is not unexpected, as these rooms are very similar in terms of RT and DRR as shown in Fig.~\ref{fig:expDataSetRts}. Using measured responses from a database according to parameter similarity might thus in fact be another viable approach for solving the room acoustic matching problem~\cite{Klein2019}. Comparing the results from part I.b (white bars in Fig.~\ref{fig:partIab}) however shows that when comparing measurements from these two rooms, distinguishability was far from perfect (a result in line with \cite{Helmholz2022}) but significantly above chance level. This suggests that the estimation is not perfect yet, as in that case, participants would recognize the differences between the rooms even when one of them used an estimated response for rendering.

Considering the more direct question (ii) of whether an estimated response conveys the room acoustics as well as a measured response, the experiment has shown that for one of the three rooms, the estimate is selected as often as a measurement from the same position. For the other two rooms, the measurement was selected significantly more often, suggesting that the method does not provide equally good results for all types of rooms yet.

The presented perceptual results are based on estimates obtained under noise-free conditions with knowledge of the true source DOA to investigate the best-case perceptual performance of the method. To offer further insights, we provide audio samples from the listening experiment and additional binaural renderings based on estimates under non-ideal conditions online.\footnote{\texttt{\url{https://facebookresearch.github.io/GlassesRoomID/}}}
Informal listening revealed that most estimates under non-ideal conditions still accurately reproduce the room acoustics, consistent with the method's observed robustness in Sec.~\ref{sec:robustness}. However, some examples show minor coloration, and in more extreme low-SNR cases or those with significant DOA offsets, slight changes in reproduced reverberation or distance occur. In scenarios with strong interference, the overall spatial impression may be compromised.


%


\section{Conclusion}\label{sec:conclusion}
We presented a method to blindly identify BRIRs from speech signals captured with a microphone array in a pair of smart glasses. In an intermediate step, the method provides multichannel RIR estimates that were used to estimate room acoustic parameters.
When using far-field speech, the method outperforms baseline estimators in both RT and DRR estimation in all considered scenarios, across different microphone array configurations and SNRs. It further reproduces the directional energy distribution captured by the multichannel RIRs similarly accurate with the pair of smart glasses as with a comparable conventional microphone array. The method is robust against inaccuracies in the assumed source DOA, deviations in the chosen block length, and interfering speech.
When the proposed method uses speech from the person wearing the smart glasses for the estimation, it delivers accurate RT estimates only in high-SNR scenarios. High-SNR scenarios are however the most realistic scenarios for this use case as the proximity of the user's mouth to the microphone array naturally ensures a high SNR.

The estimated BRIRs were further evaluated in a listening experiment. The results suggest that the estimated BRIRs often allow for a perceptually more convincing virtual source rendering than measured BRIRs from other rooms of similar size. For one of the three investigated rooms, the BRIR estimates were selected as often as a measured BRIR from the same room, showing that the estimate was as similar to the reference as the measurement. In no case was a measurement from a different room selected more often than a generated estimate, indicating that the estimates do not contain artifacts or distortions that would make participants undoubtedly choose the other given stimulus. 

So far, the method was evaluated using a single static source and a static microphone array. Future work should extend the proposed method to support adaptive online RIR estimation under more complex acoustic conditions including multiple competing sources, moving sources, and a moving microphone array. Additional processing blocks to tackle these problems may include a selection module for the pseudo reference source, voice activity detection, source tracking, and compensation for array movement.

A reference implementation and binaural audio samples from the listening experiment are provided at \texttt{\url{https://github.com/facebookresearch/GlassesRoomID}}.

\section*{Acknowledgment}
The authors thank Andy Stockton and Andrew Francl for fruitful discussions. Thomas Deppisch and Nils Meyer-Kahlen thank the RLR Audio Team at Meta for the welcoming and encouraging environment during their internships.








\bibliographystyle{IEEEtran}
\bibliography{refsTommi,refsNils}
\input{bios}

\end{document}

%% file: bios.tex
\begin{IEEEbiography}[{\includegraphics[width=1in,height=1.25in,clip,keepaspectratio]{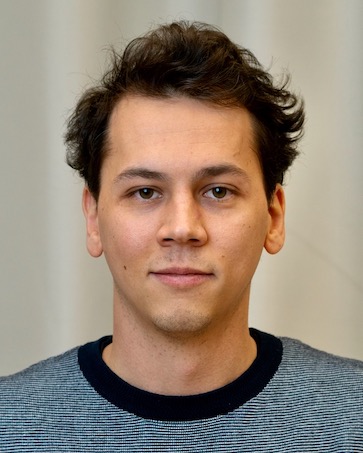}}]{Thomas Deppisch}
received the M.Sc. degree in Electrical Engineering and Audio Engineering jointly from Graz University of Technology and the University of Music and Performing Arts, Graz, Austria, in~2020. Since 2020, he has been working toward the Ph.D. degree at the Division of Applied Acoustics at Chalmers University of Technology, Gothenburg, Sweden. He was a research scientist intern at Meta Reality Labs Research in Redmond, WA, USA, in 2023. His main research interests lie in signal processing methods for the virtual reproduction of acoustic environments, including their capture, analysis, rendering, and perception.
\end{IEEEbiography}
\begin{IEEEbiography}[{\includegraphics[width=1in,height=1.25in,clip,keepaspectratio]{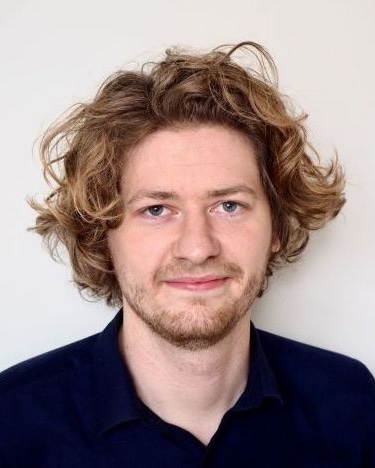}}]{Nils Meyer-Kahlen}
is a postdoctoral researcher at the Department of Information and Communications Engineering at Aalto University in Finland, where he defended his doctoral thesis in 2024.  Before joining the lab in 2019, he completed his B.Sc. and M.Sc. in Electrical Engineering and Audio Engineering at the Technical University and the University of Music and Performing Arts in Graz, Austria. In 2022 and 2024, he was a research scientist intern at Meta Reality Labs in Redmond, WA, USA. His main research interest is virtual acoustics for augmented reality from both a technological and a perceptual point of view.
\end{IEEEbiography}
\begin{IEEEbiography}[{\includegraphics[width=1in,height=1.25in,clip,keepaspectratio]{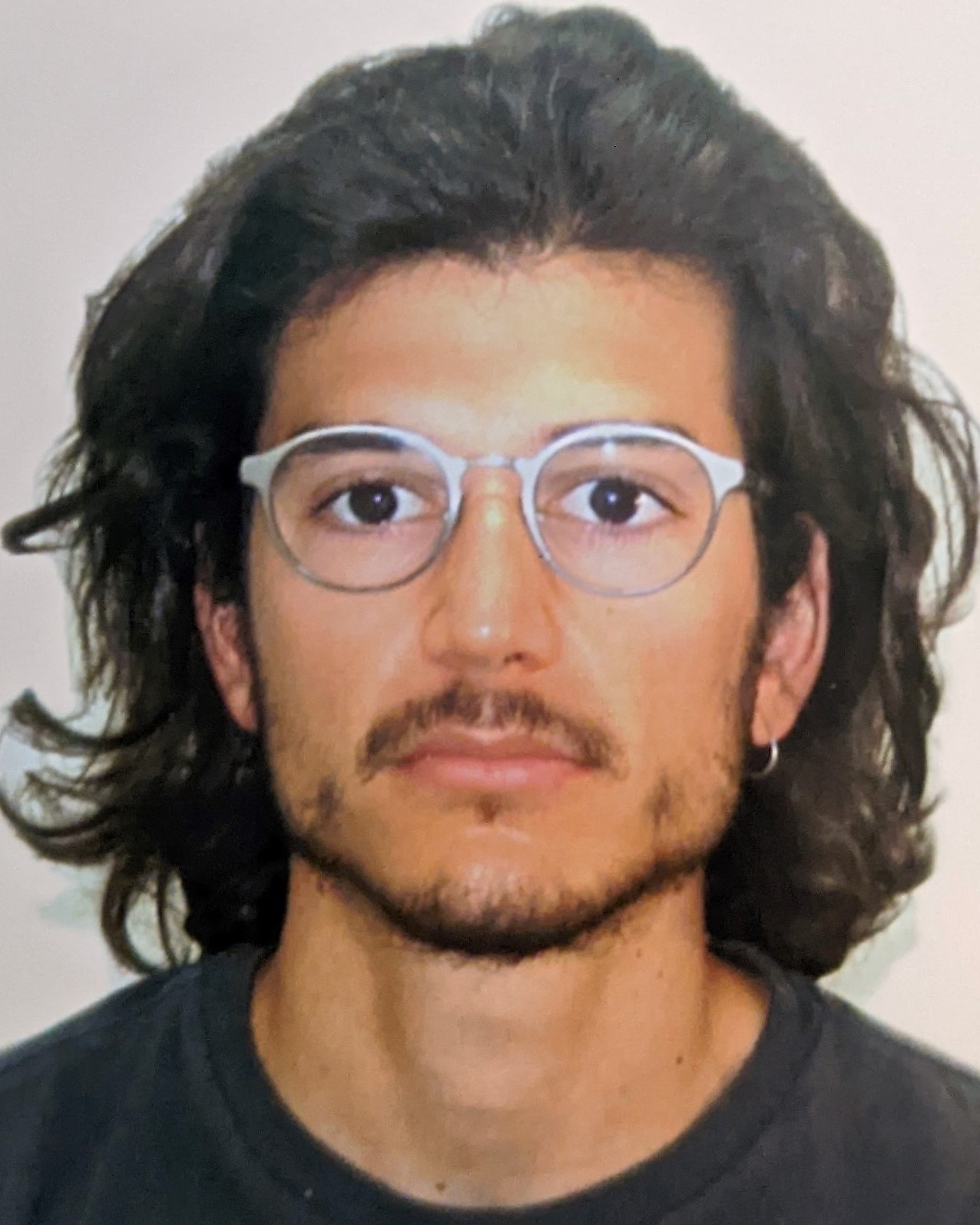}}]{Sebastià~V.~Amengual~Garí}
is currently a research scientist at Reality Labs Research working on room acoustics, spatial audio and auditory perception. He received a Diploma Degree in Telecommunications with major in Sound and Image in 2014 from the Polytechnic University of Catalonia (UPC) in 2014, completing his Master's Thesis at the Norwegian University of Science and Technology (NTNU). His doctoral work at the Detmold University of Music focused on investigating the interaction of room acoustics and live music performance using virtual acoustic environments. His research interests lie in the intersection of audio, perception and music.
\end{IEEEbiography}
\vfill